\documentclass[%
 reprint,
nofootinbib,
 amsmath,amssymb,
 aps,
]{revtex4-2}
\usepackage[colorlinks = true,urlcolor=blue,linkcolor = blue,citecolor=blue]{hyperref}
\usepackage{graphicx}
\usepackage{dcolumn}
\usepackage{bm}
\usepackage{ulem}

\usepackage{color}
\usepackage{comment}
\usepackage[export]{adjustbox}
\usepackage{soul}
\usepackage{physics}
\usepackage{tabularx}
\usepackage{multirow}
\usepackage{tikz}

\newcommand{\tim}[1]{ { \color{red} \footnotesize (\textsf{Tim}) \textsf{\textsl{#1}} } }

\newcommand{\subh}[1]{{\color{blue}#1}}

\newcommand{\Amir}[1]{{\color{green}#1}}

\begin{document}

\preprint{}

\title{Measurement-induced phase transitions in the toric code}%
\author{Amir-Reza Negari}
\author{Subhayan Sahu}
\author{Timothy H. Hsieh}%

\affiliation{%
 Perimeter Institute for Theoretical Physics, Waterloo, Ontario N2L 2Y5, Canada}%

\affiliation{%
Department of Physics and Astronomy, University of Waterloo, Waterloo, Ontario N2L 3G1, Canada}%
 
\begin{abstract}

We show how distinct phases of matter can be generated by performing random single-qubit measurements on a subsystem of toric code. Using a parton construction, such measurements map to random Gaussian tensor networks, and in particular, random Pauli measurements map to a classical loop model in which watermelon correlators precisely determine measurement-induced entanglement.  Measuring all but a 1d boundary of qubits realizes hybrid circuits involving unitary gates and projective measurements in 1+1 dimensions.  We find that varying the probabilities of different Pauli measurements can drive transitions in the un-measured boundary between phases with different orders and entanglement scaling, corresponding to short and long loop phases in the classical model.  Furthermore, by utilizing single-site boundary unitaries conditioned on the bulk measurement outcomes, we generate mixed state ordered phases and transitions that can be experimentally diagnosed via linear observables. 
This demonstrates how parton constructions provide a natural framework for measurement-based quantum computing setups to produce and manipulate phases of matter.

\end{abstract}
\maketitle

\tableofcontents

\section{Introduction}

Investigating the quantum phases of matter that can be dynamically generated in a quantum processor using measurement, classical feedback, and local unitaries  has been a fruitful area of research. There has been significant interest in using such hybrid circuits to manipulate entanglement patterns, starting with the observation of a measurement-induced entanglement transition between volume law to area law as the frequency of measurements is varied~\cite{aharonov2000quantum,li2018quantum,skinner2019measurement,li2019measurement,gullans2020dynamical,zabalo2020critical,Li2021conformal,jian2020measurement,bao2020theory} (for a review, see~\cite{FisherReview2023,Potter2022}). 
Even without any unitary gates, random measurements of multi-site operators can lead to not only various entanglement patterns but also distinct long-range orders, such as symmetry-protected topological order, spin-glass order, and intrinsic topological order ~\cite{Sang_2020,Lavasani_2020, lavasani2022monitored,sriram2022topology}. These orders can undergo phase transitions by adjusting the probabilities of competing measurements. 

A different context in which measurements take center stage is measurement-based quantum computation (MBQC)~\cite{raussendorf_measurement-based_2003}. The MBQC approach involves starting with an entangled ``resource state'', such as the 2D cluster state, and sequentially performing single site measurements on the majority of the qubits, where the measurement basis can depend on the outcomes of previous measurements. This results in the remaining un-measured qubits being directed towards a specific entangled state that encodes the outcome of a deterministic quantum computation.  For example, measurements on a 2d resource state effectively realize a computation on the 1d boundary of the system, and the other dimension corresponds to the ``time'' direction of the computation.  

In this work, we ask the question: starting with an entangled resource state, can MBQC-type protocols lead to robust quantum phases of matter and transitions between them? We explore this question for the toric code ground state, which is an exactly solvable model of $\mathbb{Z}_{2}$ topological order in two dimensions ~\cite{Kitaev1997}. We find that by tuning the probabilities of measuring single-site Pauli $X$,$Y$, or $Z$ in the toric code bulk, we can realize distinct phases in an un-measured boundary. These measurement-induced phases are characterized by the presence or absence of spin-glass order parameters and their entanglement scaling (area law vs. logarithmic scaling).  
As is the case for MBQC, the bulk measurements in the toric code effectively realize dynamics for a one-dimensional boundary, and in this case the effective dynamics are that of a hybrid circuit involving both unitaries and measurements.  Thanks to the underlying entanglement of the toric code state, only single-qubit measurements are required to effectively realize non-trivial hybrid circuits involving two-qubit operations.  

\begin{figure}[h]                       
\includegraphics[width=0.9\columnwidth]{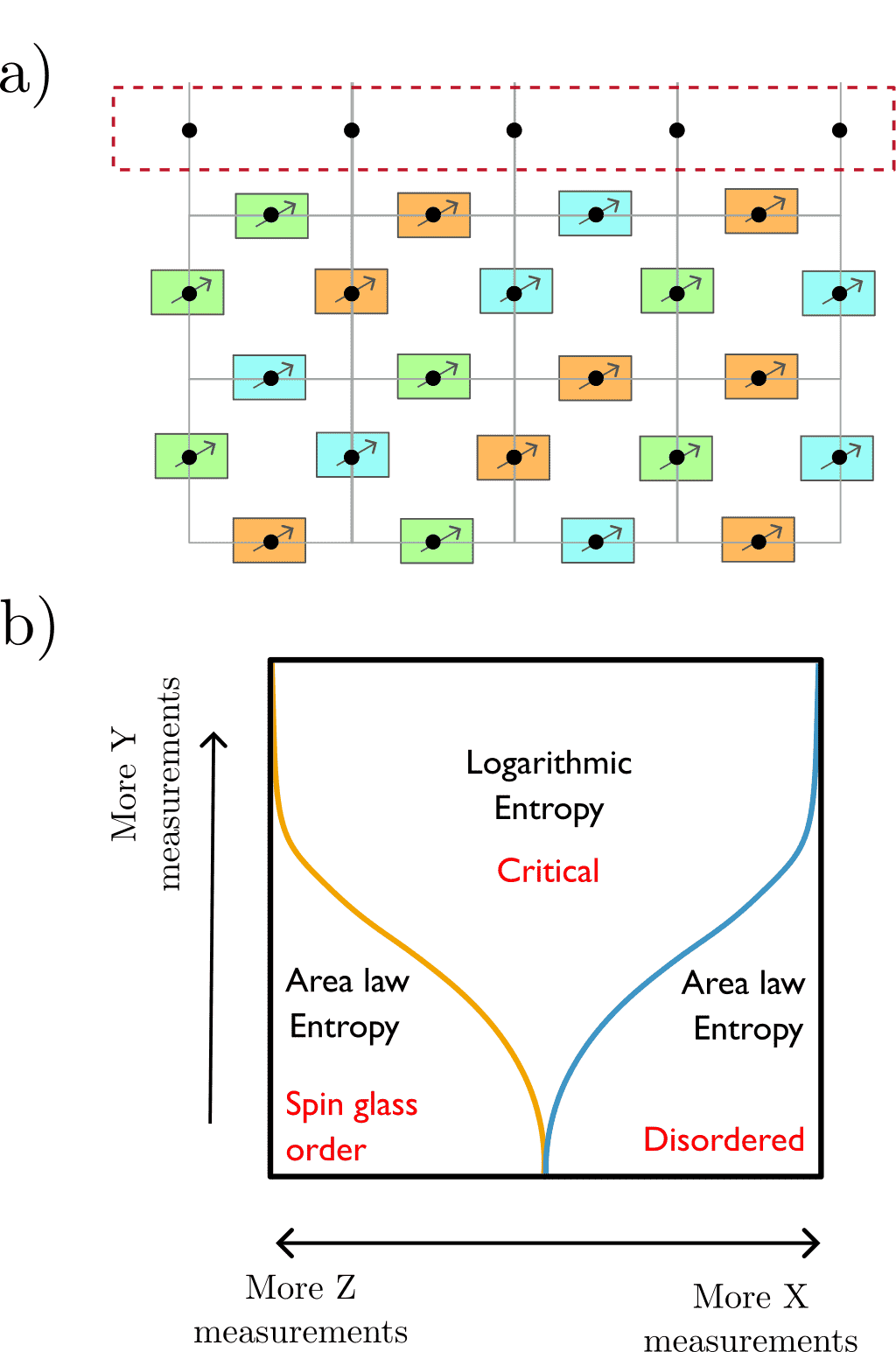}
    \caption{(a) The bulk of a toric code ground state is measured in random Pauli bases (denoted by different colors), which induces correlations between the unmeasured 1d boundary qubits (in dashed box). (b) Depending on the relative frequencies of different Pauli measurements, the 1d boundary can have different entanglement scaling (area law vs. logarithmic scaling) and also different orders (spin-glass vs. paramagnetic).  Such phases and transitions are analyzed by mapping to a 2d classical loop model.}
    \label{fig:adv}
\end{figure}

Liu et al.\cite{Liu2022} performed a related study on the 2D cluster state, an MBQC resource state 
 which enables universal quantum computation, and discovered an entanglement transition from area to volume law in boundary qubits induced by measuring the bulk qubits.  In contrast, in our setup with toric code, we find transitions in both entanglement (albeit without volume law) and other order parameters.  A recent study \cite{yang} also considered an MBQC setup on the 2d cluster state and found evidence of distinct area law entanglement phases on the 1d boundary.  One advantage of our setup is that we can understand such transitions analytically by relating the entanglement properties of the qubits to certain correlation functions of a corresponding 2d classical loop model with crossings. Such a model has short and long loop phases, which exactly correspond to the area law and the logarithmic scaling of entanglement in the 1d boundary.  The summary of these results is presented in Fig.~\ref{fig:adv}.  We also note that \cite{behrends2022surface} established a very different mapping between toric code error correction in the presence of both incoherent and coherent errors and $(1 + 1)d$ free-fermion circuits.

As is the case with hybrid circuits without feedback (measurement outcomes are not used to inform future operations), the transition between quantum phases is only apparent in quantities nonlinear in the ensemble of quantum trajectories.  In our examples, these nonlinear quantities can be spin-glass order parameters or entanglement measures.   
However, the spin glass order can be converted into ferromagnetic order (a linear observable) via feedback: we show that one layer of single-qubit unitaries, conditioned on the bulk measurement outcomes, can be applied to the boundary state to ensure that the resulting density matrix averaged over trajectories has long-range order that is observable. This constitutes a nontrivial quantum channel on a 2d array producing a long-range entangled mixed state in 1d; as in \cite{lu2023mixedstate}, it relies on measurement and unitary feedback, though the resulting mixed state is likely difficult to generate using only operations on the 1d system. 

This setup can be readily generalized from Pauli $X,Y,Z$ measurement to arbitrary single-site projective measurements. We find that such measurements in $2+0$d map in general to Gaussian fermionic hybrid circuits in $1+1$d. This mapping allows us to import the results about entanglement phases generated by such circuits (for example ~\cite{jian2022criticality,jian2023measurementinduced,Merritt_2023}) onto the measurement-induced entanglement on the boundary state of the toric code. Even in the general on-site measurement setup, the phases with area law and logarithmic scaling of entanglement persist, albeit with distinct phase boundaries and transitions.


The structure of the paper is as follows: in section~\ref{sec:setup}, we introduce the measurement setup for toric code. We then map the stabilizer configurations after measurements to the completely packed loop model with crossing (CPLC) and summarize relevant results in section~\ref{sec:loop_mapping}. In section~\ref{sec:MIE}, we relate specific order parameters in the loop model to entanglement induced by measurements between different regions. In section~\ref{sec:bulk_meas}, we explain how the mapping leads to distinct entanglement patterns in un-measured boundary qubits and we demonstrate how the setup can be mapped to a 1+1D hybrid circuit. In section~\ref{sec:order}, we show how the presence or absence of a certain spin glass order distinguishes the two phases. Furthermore, we describe a simple adaptive protocol that modifies the boundary state and enables identification of the two phases based on linear order parameters of the state. In section \ref{sec:beyond_cplc} we analyze general single-qubit measurements (beyond Pauli) on the toric code and map the resulting states to Gaussian tensor networks and Gaussian hybrid circuits. This section also contains a tensor network representation of the toric code ground state via parton construction, which may be of independent interest. Finally, in section~\ref{sec:discussion} we conclude with a discussion of our results, including relations to the underlying sign structure and MBQC universality of the resource state.

\section{Setup}\label{sec:setup}

\subsection{Toric code/Plaquette model}

\begin{figure}[h]                       
\includegraphics[width=0.9\columnwidth]{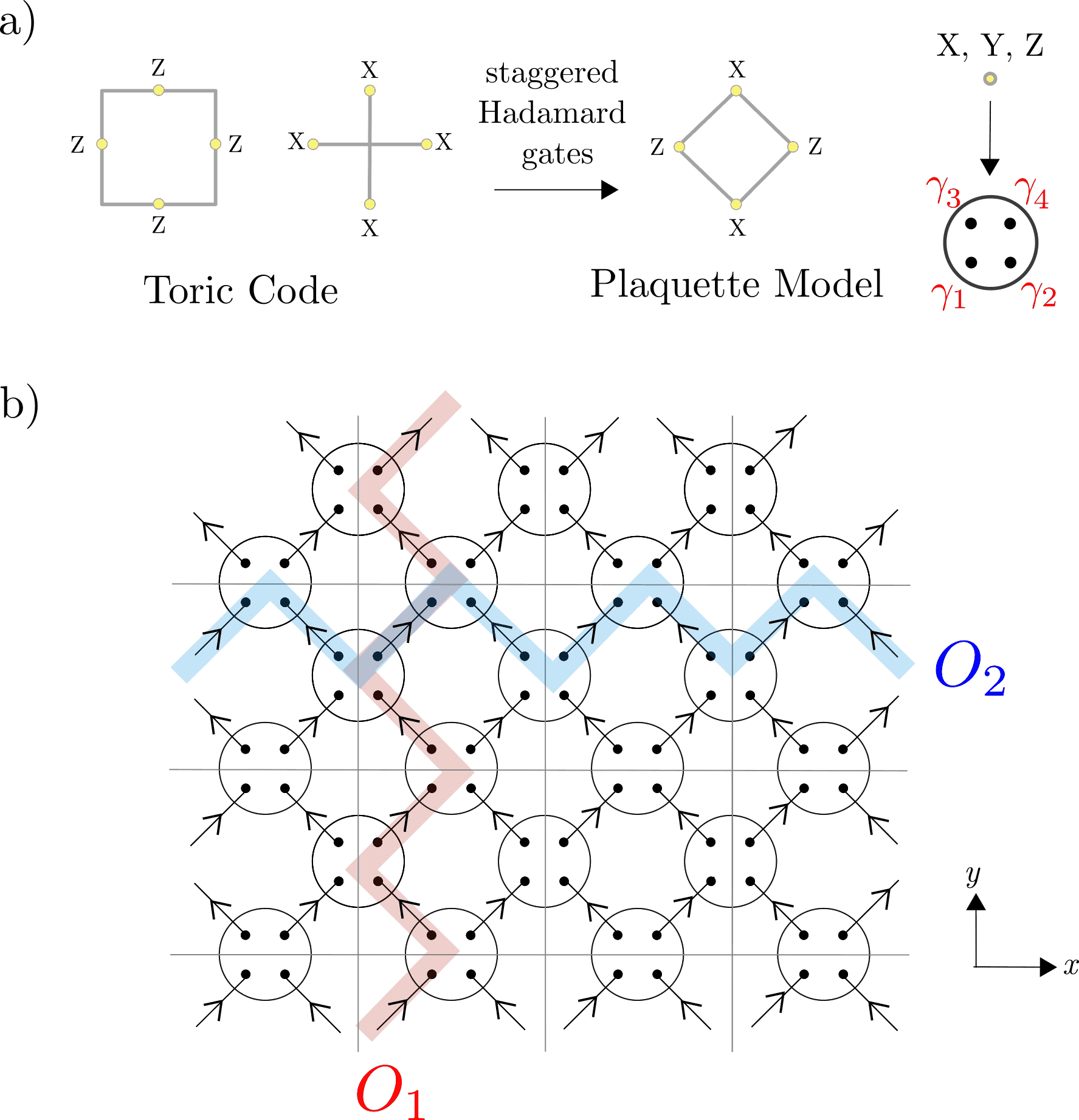}
    \caption{(a) Toric code stabilizers can be converted to Wen plaquette stabilizers via staggered Hadamard gates.  The Wen plaquette admits a parton construction in which each qubit is split into four Majorana fermions subject to a constraint (represented by circle in right subfigure). (b) A ground state of the plaquette model can be constructed by projecting a free fermion state consisting of Majorana dimers into the physical qubit Hilbert space.}
    \label{fig:Model}
\end{figure}

The toric code is a lattice model of spin-1/2 degrees of freedom on the edges of a square lattice \cite{Kitaev1997} which consists of commuting terms in its Hamiltonian called stabilizer operators. The toric code has two types of stabilizer operators: star (s) and plaquette (p) operators, 

\begin{equation}
    H_{T} = -\sum_s \prod_{j\in s} X_j -\sum_p \prod_{j\in p} Z_j
\end{equation}
A closely related model is Wen's 2D plaquette model \cite{Wen2003}, where the spin-1/2 degrees of freedom are located at the vertices of a square lattice, and the Hamiltonian consists of only one type of 4-body stabilizer for every star $s$ and plaquette $p$, on the $45^{\circ}$-rotated lattice (see Fig.~\ref{fig:Model}a):  
\begin{equation}
    H_{W} = -\sum_{a \in p,s} X_{a+\hat{y}} Z_{a+\hat{x}} X_{a-\hat{y}} Z_{a-\hat{x}}
\end{equation}

These two models can be transformed into each other using a single layer of local Hadamard gates arranged on one (say, B) of the two sub-lattices (A and B) of the square lattice in the plaquette model. Sublattice A represents spins on vertical edges, while sublattice B represents spins on horizontal edges of toric code in Fig ~\ref{fig:Model}b. These gates interchange $X \leftrightarrow Z$ on the B sub-lattice, which interchange the plaquette model and toric code, as depicted in Fig.~\ref{fig:Model}a. On a torus defined by identifying the boundaries along $x,y$ directions as marked in Fig.~\ref{fig:Model}b, the toric code has four degenerate ground states, labeled by $\pm 1$ eigenvalues of the logical operators $O'_1, O'_2$. These logical operators can be obtained by applying the previously mentioned Hadamard gates to Wen's logical operators $O_1$ and $O_2$, which consist of strings of Pauli-Z and Pauli-X operators as depicted in Fig.~\ref{fig:Model}b.

Any eigenstate $\ket{G}$ of the toric code admits an exact free fermion parton construction \cite{Wen2003,Kitaev_2006} defined as follows. The Hilbert space of each qubit on site $i$ can be enlarged into that of four Majorana fermions $\gamma_{i,s} = \{\gamma_{i,1},\gamma_{i,2},\gamma_{i,3},\gamma_{i,4}\}$ along the edges connected to $i$, followed by a projection onto the original qubit Hilbert space. Consider the free fermion state $\ket{\psi}_{\text{free}}$ such that $i\gamma_{i,s} \gamma_{j,s'}=1$ when $(i,s),(j,s')$ are on the same edge.  To return the qubit Hilbert space, we must project the Majorana state to the $+1$ sector of the operator $D_j = \gamma_{j,1} \gamma_{j,2} \gamma_{j,3} \gamma_{j,4}$: 

\begin{equation}
    |G\rangle = \prod_j\left(\frac{1+D_j}{2}\right)\ket{\psi}_{\text{free}}
    \label{eq:GenVarState}.
\end{equation}

Note that the initial free fermion state of the two Majorana modes on neighboring vertices can be oriented in two different ways, depending on whether we take the $+1$ eigenstate of $\pm i\gamma_{i,s}\gamma_{j,s^{\prime}}$. Different orientations (which are marked by $s\rightarrow s^{\prime}$ to indicate $+i\gamma_{s}\gamma_{s^{\prime}}$ in Fig.~\ref{fig:Model}b) determine the particular eigenstate up to a global phase. In particular, the groundspace of the toric code on a torus is 4 dimensional; one representative ground state is described in our convention by the orientation shown in Fig.~\ref{fig:Model}b, where the same orientation is taken along all $45^{\circ}$ lattice lines. 
Different logical sectors (choices of $O_1,O_2=\pm 1$) of the plaquette model ground space can be represented by flipping all the orientations of the links $ij$ along the non-trivial loops. We will focus our attention on the ground state defined by the orientation shown in Fig.~\ref{fig:Model}b, which corresponds to $O_{1}, O_{2} = +1$.

\subsection{Measurement setup}

First we consider the case of measuring a subset of qubits $M$ in the toric code in either $Z$, $Y$, or $X$ bases, with respective probabilities $(1-q)(1-p)$, $p$, and $q(1-p)$, which we call the $(p,q)$ measurement protocol. Our objective is to analyze the entanglement structure and order in the remaining (un-measured) qubits $M^c$ after the measurements on $M$ are performed. The target quantities of interest are averaged over all realizations of both measurement configurations and their outcomes. In a later section, we will generalize to the case of measuring along any direction in the Bloch sphere.

Due to the equivalence between the toric code and plaquette models via a Hadamard transformation on one sub-lattice, the $(p,q)$ scheme for toric code is equivalent to the  $(p,q)$ scheme on A sublattice and $(p,1-q)$ scheme on B sublattice for the plaquette model.

In the plaquette model, Pauli operators on site $j$ correspond to Majorana fermion bilinear operators
\begin{equation}
\begin{split}
    X_{j}= i\gamma_{j,1}\gamma_{j,2} = i\gamma_{j,4}\gamma_{j,3}  \\ Y_{j}=  i\gamma_{j,2}\gamma_{j,3} = i\gamma_{j,4}\gamma_{j,1} \\ Z_{j}= i\gamma_{j,1}\gamma_{j,3} = i\gamma_{j,2}\gamma_{j,4}, 
\end{split}
\end{equation}
where the right equalities follow from the physical Hilbert space condition ($D_j=1$).


\subsection{Stabilizers and measurement}\label{sec:setup_meas}
We first provide a brief overview of the Majorana stabilizer formalism specialized to our setting.  The set of stabilizer generators  $\mathcal{G}$ is  a set of products of Majorana fermions which are independent and mutually commute with each other. This set generates the stabilizer group $\mathcal{S}$. In a Hilbert space of dimension $2^N$, a set $\mathcal{G}$ with exactly $N$ generators uniquely defines the common eigenvector $\ket{\psi}$ of any operator generated by $\mathcal{G}$, such that $s\ket{\psi} = \ket{\psi} \forall s \in \mathcal{S}$.

If we measure the state $\ket{\psi}$  with an operator $P$ which is a product of Majorana fermions, the resulting state is still a Majorana stabilizer state and can be updated efficiently~\cite{gottesman1998heisenberg,aaronson2004improved}.  There are two cases to consider. If $P$ commutes with all the stabilizer generators $g\in \mathcal{G}$, the measurement will not have any effect on the state, and the measurement outcome can be inferred from the sign of the operator in $\mathcal{S}$, i.e., whether $\pm P \in \mathcal{S}$. If $P$ anti-commutes with some of the stabilizer generators, the measurement outcome $\pm 1$ with equal probability. We also have to modify the set of generators $\mathcal{G}$ - first we select one of the anti-commuting generators, denoted as $g_0$, and multiply $g_0$ with the remaining anti-commuting generators. Next, we replace $g_{0}$ in $\mathcal{G}$ by either $\pm P$ depending on the measurement outcome, so that the new stabilizer set becomes $\{\pm P\}\cup\{ g_{0}g_{i}| \ \forall i\neq 0,g_{i} \text{ anti-commutes with } P\}\cup \{ g_{i}| \ g_{i} \text{ commutes with } P\}$. 

The stabilizer formalism offers a way to confirm that the ground state $\ket{G}$ is the projected free fermion state $|\psi\rangle_{\text{free}}$ , stabilized by two-point Majorana fermion operators: $i\gamma_{j,s}\gamma_{i,s'}$. The action of projection operators on the stabilizers exclusively modifies adjacent Majorana pairs by multiplying them together to form string operators. By applying all the projections, these string operators eventually become closed loops, forming Majorana loop operators which act as stabilizers for the ground state in Wen's model. Our goal is to measure Pauli operators corresponding to different two-point Majorana operators on every site.  Since these are physical qubit operators, they commute with the projection operator, and hence we can first consider their effect on the free fermion state before applying the projection operator at the end.

\begin{figure}[ht]    \includegraphics[width=0.9\columnwidth]{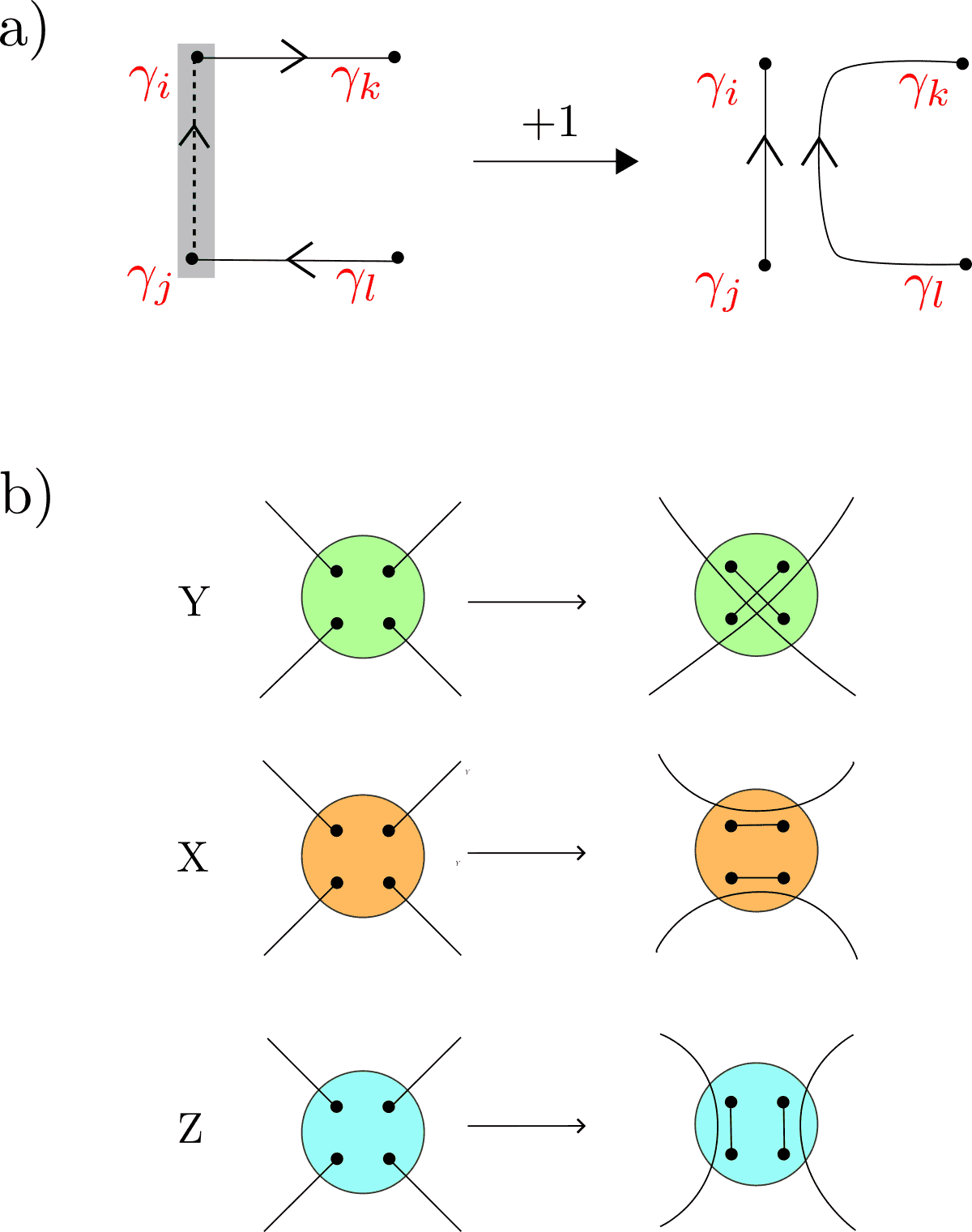}
    \caption{(a) The update process for measuring the Majorana bilinear $i\gamma_j \gamma_i$ for a free fermion state which is a tensor product of Majorana dimers. The $+1$ refers to the measurement outcome, which sets the arrow direction in the final state. If the measurement outcome is $-1$ the final arrows need to be reversed (Detailed analysis in part II section C). (b) Updated pairings from measuring $X$, $Y$, and $Z$ Pauli operators on each site of the plaquette model. The updated pairings need to be tiled to form the global dimer state. We have neglected the sign tracking of the dimer pairs, which depends on the measurement outcomes.}
    \label{fig:Majorana_parings}
\end{figure}

We graphically track the free fermion state updates by  
connecting Majorana fermions with a line when they form a stabilizer operator together.  When $i\gamma_j\gamma_i$ is measured, there are two possible outcomes: (a) If there is already a connection between $\gamma_i$ and $\gamma_j$ in the initial state, no further updates are required, and, (b) If these two Majorana fermions are connected to other Majoranas (e.g., $\gamma_i$ is connected to $\gamma_k$ and $\gamma_l$ is connected to $\gamma_j$), the update will connect $\gamma_j$ to $\gamma_i$, and the other Majoranas will be connected accordingly (e.g., $\gamma_l$ to $\gamma_k$), as shown in Fig.~\ref{fig:Majorana_parings}a. 


The signs of stabilizers and measurement outcomes can be tracked and updated by using arrows on Majorana pairings, as illustrated in Fig.~\ref{fig:Majorana_parings}a.  However, the signs will not be important when computing entanglement quantities or spin glass order parameters, in the case of $X,Y,Z$, i.e. stabilizer, measurements.
In the next sections, we will suppress the arrow notation for signs and return to the task of sign-tracking when discussing the linear order parameter in Sec.~\ref{sec:order} and on-site measurements in general directions in Sec.~\ref{sec:beyond_cplc}.




\section{Completely packed loop model}\label{sec:loop_mapping}

Measuring Pauli operators on each site generates three different patterns of pairings (Fig.~\ref{fig:Majorana_parings}b), and measuring all qubits tiles these patterns and results in different configurations of loops on a square lattice. On the two different sub-lattices of the square lattice, the factors $q$ and $1-q$ must be swapped, to reflect the staggered measurement scheme of the plaquette model.


Consider a configuration of measurements or tilings, with the  $N_{x}, N_{y}, N_{z}$ number of $X, Y, Z$ measurements performed. Such a configuration has probability $W_{\mathcal{C}} = p^{N_y} [(1-p)q]^{N_x} [(1-p)(1-q)]^{N_z}$ which leads to the partition function $Z =\sum_{\mathcal{C}}W_{\mathcal{C}}$. The model and partition function are known as the completely packed loop model with crossings (CPLC), whose properties have been extensively studied in \cite{Nahum_2013}. We will now review its important properties relevant to the questions addressed in this work.

\begin{figure}[h]
    \includegraphics[width=1\columnwidth]{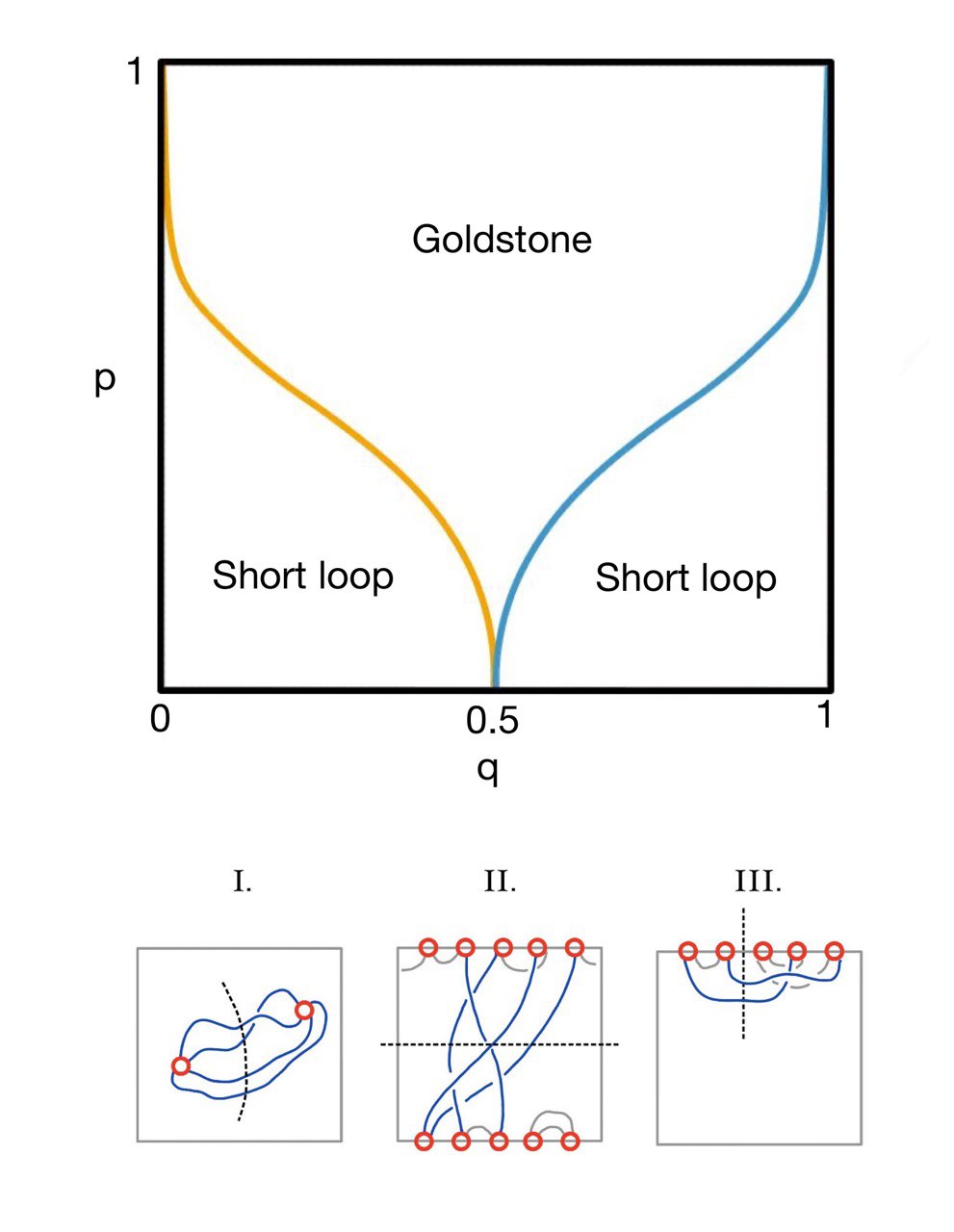}
    
    \caption{The schematic phase diagram of the completely packed loop model with crossings (CPLC)~\cite{Nahum_2013}.   The diagram distinguishes between different phases based on three order parameters, highlighted in the panel below. These order parameters correspond to loop configurations that connect points marked by red dots. The first order parameter (I) quantifies the probability of four distinct strands connecting two red points on a torus, referred to as the ``watermelon correlation function." The second order parameter (II) measures the expected number of strands connecting the top and bottom boundaries on a cylinder, known as the ``spanning number." The third order parameter (III) measures the expected number of strands connecting two partitions of the top boundary on a cylinder with a fixed boundary condition at the bottom. All three quantities govern measurement-induced entanglement in the toric code state.}
    \label{fig:phase_diagram}
\end{figure}

In \cite{Nahum_2013} the authors found that the phase diagram consists of a short loop phase and a long loop ``Goldstone'' phase, which are separated by a phase transition (see Fig. \ref{fig:phase_diagram}). This model can be described by the replica limit $n\to 1$ of a $\mathbb{Z}_{2}$ lattice gauge theory coupled to $O(n)$ matter field. Its continuum description is a sigma model, which is massive in the short loop regime and massless in the Goldstone phase~\cite{Nahum_2013}.

We focus on two order parameters which distinguish the phases. First, we consider the \textit{watermelon correlation functions} $G_k(i,j)$, which denote the probability that $k$ distinct strands connect points $i$ and $j$, where $k$ is even for the CPLC model. For instance, 
$G_4$ is the probability that two nodes are connected by four distinct strands. Using renormalization group (RG) techniques on the sigma model, \cite{Nahum_2013} found that in the Goldstone phase 
\begin{eqnarray}\label{eq:watermelon}
G_k(i,j) \sim
\frac{C_0}{\ln\left(d_{ij}/r_0\right)^{k(k-1)}} 
\end{eqnarray}
where $d_{ij}$ is the distance between $i,j$ and $C_{0}, r_{0}$ are non-universal constants . In the short-loop phases on the other hand, the watermelon correlators decay as $G_{k}(i,j) \sim e^{-d_{ij}/\xi}$, with correlation length $\xi$.

Next, we consider the \textit{spanning number} defined for a CPLC model on a cylinder, with two circular open boundaries. The spanning number counts the number of strands that connect the upper and lower boundaries. \cite{Nahum_2013} found that in the Goldstone phase, the average spanning number scales with system size $L$ as 
\begin{eqnarray}\label{eq:spanning_number}
    n_{s} \approx \frac{1}{2\pi} \left(\ln \frac{L}{L_{0}}+\ln\ln \frac{L}{L_{0}}\right),
\end{eqnarray}
whereas it asymptotes to $0$ in the short loop phase.

To explore the entanglement properties of the toric code after measurements and their connection to the phase transitions in the loop model, we need to leave some qubits un-measured as measuring all qubits results in a trivial pure product state. Three scenarios are considered (see schematic description in Fig.~\ref{fig:phase_diagram}): 

(I) \textit{Measuring all but two qubits in the bulk.} In Sec.~\ref{sec:MIE_qubit} we show that the entanglement induced between the two un-measured qubits is directly related to the watermelon correlation function.

(II) \textit{Measuring all but two boundaries.} In Sec.~\ref{sec:MIE_two_boundary}, we observe that the induced entanglement between the two boundaries of the cylinder is directly related to ``spanning number" order parameter discussed in this section. Accordingly, in the short loop phase, the entanglement is asymptotically zero, while in the Goldstone phase, it exhibits logarithmic scaling with the system size.

(III) \textit{Measuring all but a single boundary.} We show in Sec.~\ref{sec:bulk_meas} that in this case the entanglement between contiguous bipartitions of the un-measured boundary exhibit a phase transition between area law and logarithmic law, reflecting the underlying loop model configurations.

\section{Measurement-induced entanglement between two distant regions}\label{sec:MIE}

In this section, we establish a connection between the average measurement-induced entanglement (MIE) of two distant unmeasured regions and various order parameters within the CPLC model. The MIE has been related to the underlying sign structure of the measured wavefunction in \cite{Lin_2023}, and we will comment more on this in the concluding discussion.  


\subsection{MIE between two un-measured qubits}\label{sec:MIE_qubit}

We first demonstrate that the measurement-induced entanglement (MIE) between two un-measured qubits at sites $i$ and $j$ is equivalent to the watermelon correlation $G_4(i,j)$.

Recall that measuring a qubit specifies a given pairing for the four Majoranas associated with the qubit.  After all pairings at all sites except for $i,j$ are specified, we must implement the projection operators $D_{k}$ on every site, as in Eq.~\ref{eq:GenVarState}. Crucially, any \textit{closed loop Majorana stabilizer commutes with the projection operator} and is thus shared by both the free-fermion and the projected state in Eq.~\ref{eq:GenVarState}. However, if two qubits are left un-measured, then some Majorana stabilizers may be open strands ending at the un-measured sites. In this case the projection operator has a significant effect on the final stabilizers and hence the entanglement between the unmeasured qubits.  

\begin{figure}[h]
    \centering
    \includegraphics[width = 0.9\columnwidth]{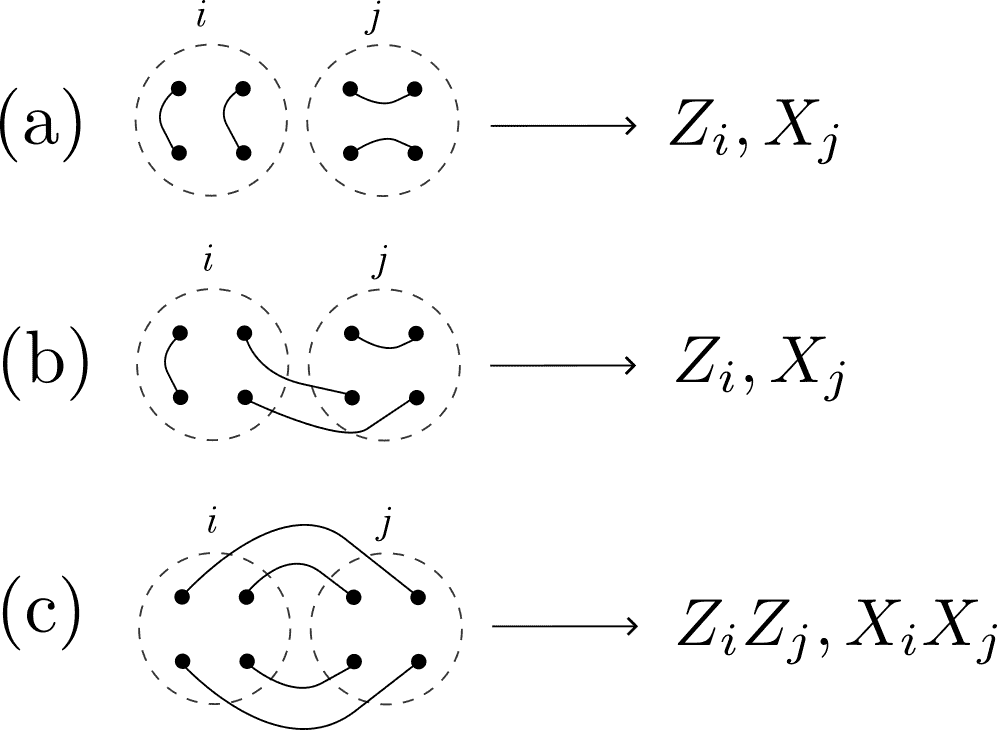}
    \caption{Three possible pairings of unmeasured qubits $i,j$ after all other qubits are measured.  Left: stabilizer strands prior to physical qubit Hilbert space projections. Right side: stabilizers after projections, in canonical form. Note, in writing the stabilizers as Pauli strings, we have assumed that the sites $i,j$ are in the same sub-lattice.
    Configurations $(a,b)$ are exclusively supported on $i$ or $j$ and do not contribute entanglement, while configuration $c$ contributes one bit of entanglement.}
    \label{fig:mie_2}
\end{figure}

To compute the measurement-induced entanglement, we analyze the three ways (Fig.~\ref{fig:mie_2}) in which Majorana stabilizer strands terminate at the two vertices $i, \ j$. (Any closed loop not coincident with $i$ and $j$ will not contribute any entanglement.)  Denote a stabilizer strand connecting Majoranas $\gamma_{i,s}$ and $\gamma_{i^{\prime},s^{\prime}}$ as $(i_{s}i^{\prime}_{s^{\prime}})$. We suppress the sign information of the stabilizer in this notation.  The three classes of configurations are 

(a) Each strand ends on Majoranas on the same vertex, i.e. we have 2 $(i_{s}i_{s^{\prime}})$ and 2 $(j_{s},j_{s^{\prime}})$ pairings.

(b) Two strands end on the same vertex and two strands end on different vertices, i.e. there are 1 $(i_{s}i_{s^{\prime}})$, 2 $(i_{s},j_{s^{\prime}})$, and 1 $(j_{s},j_{s^{\prime}})$ pairings.

(c) All four strands terminate in different vertices, i.e. there 4 $(i_{s},j_{s^{\prime}})$ pairings.

Once we impose the local projection operators $D_{i} = (i_{1}i_{2}i_{3}i_{4}), \ D_{j} = (j_{1}j_{2}j_{3}j_{4})$, the stabilizer generators need to be updated. Furthermore, to compute the entanglement between the two unmeasured qubits, we choose a canonical gauge for the stabilizers ~\cite{Nahum_2017,li2019measurement} for a given bipartition, stabilizer generators restricted to either subsystems are independent. In this canonical form, the entanglement across a bipartition is proportional to the number of stabilizers shared between both parties.

We show examples of the stabilizer update into its canonical form in Fig.~\ref{fig:mie_2}. In the right column in Fig.~\ref{fig:mie_2}, we show the stabilizer generators obtained from these patterns of strands, set in their canonical form such that the stabilizer generators restricted to either $i$ or $j$ are independent. As can be seen, the number of such independent \textit{connecting} stabilizers are $0,0,$ and $2$ respectively, in the three types of Majorana pairings $a,b,$ and $c$.  
Thus, only configuration $c$ contributes one bit of entanglement, and the average measurement-induced entanglement (MIE) generated between any two un-measured qubits on $i$ and $j$ is exactly given by the probability of 4 distinct loops connecting $i$ and $j$ in the CPLC model, i.e. the watermelon correlation function defined in the previous section,
\begin{eqnarray}
    \langle S_{\text{MIE}}(i,j) \rangle = (G^{\text{CPLC}}_{4}(i,j) )\ln 2.
\end{eqnarray}

Hence it follows from the results from~\cite{Nahum_2013} as quoted in Eq.~\ref{eq:watermelon} that the averaged MIE is long-ranged in the Goldstone phase and short-ranged in the short loop phase.

\begin{figure}[t]
    \centering
    \includegraphics[width = 0.8\columnwidth]{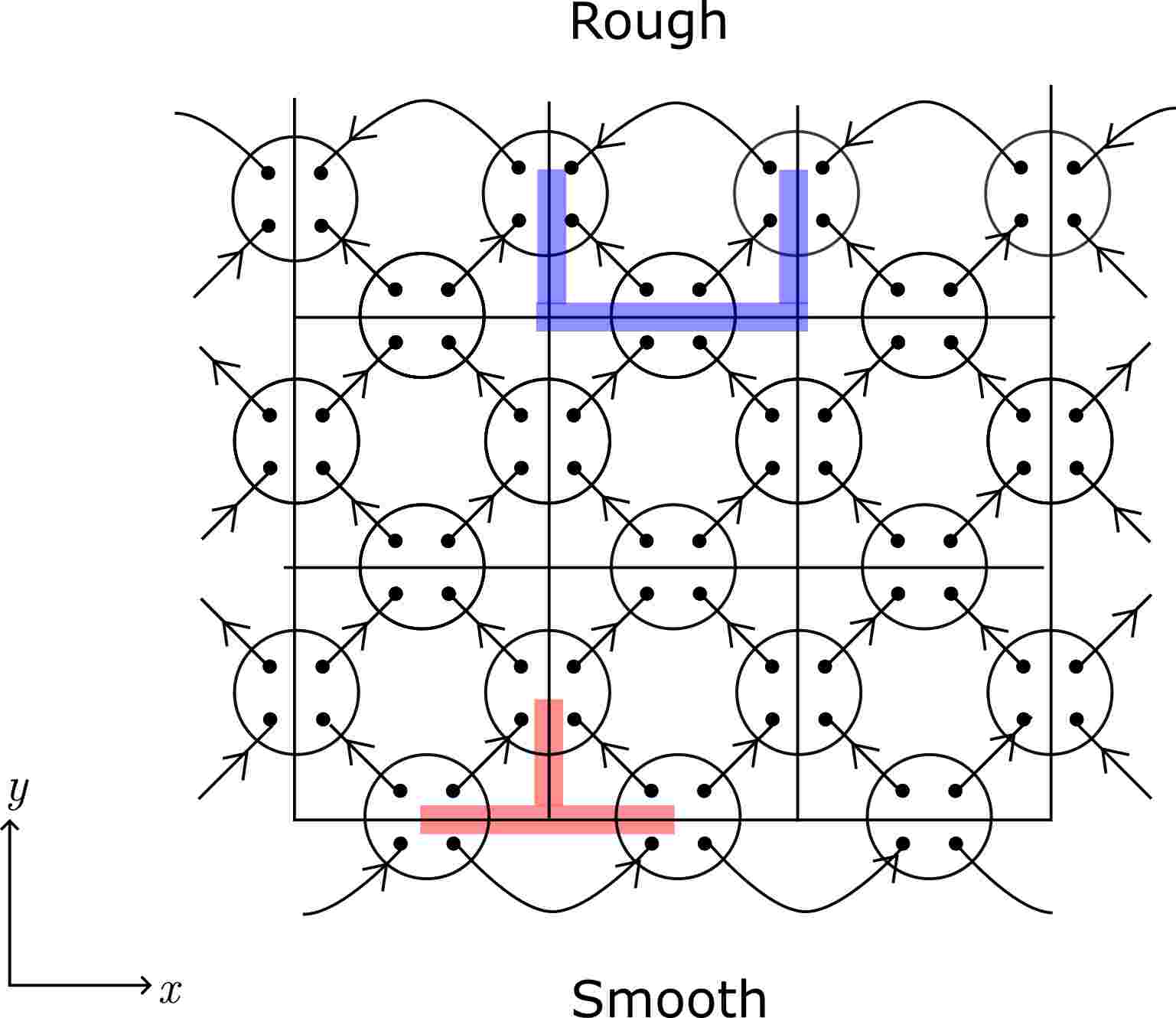}
    \caption{Boundary conditions of the parton groundstate of toric code on cylinder. The rough and smooth boundary conditions on two open edges of the cylinder are shown, along with the respective truncated stabilizers on the boundaries. The $x$ direction is taken to have periodic boundary condition.}
    \label{fig:surface_code}
\end{figure}

\subsection{MIE between two un-measured boundaries}\label{sec:MIE_two_boundary}
Now we consider the toric code on a cylinder and explore the effects of bulk measurements on the boundary chains of qubits; in particular, we focus on the scenario where both circular boundaries of the toric code are left unmeasured. For the purposes of this section, the exact boundary conditions don't matter, so we defer a discussion on the exact boundary conditions and the exact mapping of this scenario to the loop model to the next section. Here we just quote the final result, that under the Majorana loop mapping, the average entanglement between these two boundaries can be directly mapped to the `spanning number' in the loop model, as illustrated in Figure \ref{fig:phase_diagram}II. 

This can already be motivated from discussions in the earlier sub-section, where we showed that the entanglement between the two remote regions of the toric code corresponds to open strands connecting the regions in the CPLC model. However, we must also transform the stabilizers to their canonical forms in order to directly count their entanglement contribution. In Appendix IB~\cite{supp_mat} we show that in this geometry, if there are $n\geq 2$ such strands connecting the top and bottom boundaries in the loop model, we get $n-2$ independent Majorana stabilizer generators in their canonical form connecting the top and bottom boundaries. The average $n$ is just the spanning number of the loop model, so we get the following correspondence,
\begin{eqnarray}
    \langle S_{\text{MIE}} \rangle = \frac{(n_{s}-2)\ln{2}}{2}.
\end{eqnarray}

This correspondence holds true only for $n_{s}\geq 2$, otherwise, we have $S_{\text{MIE}} = 0$. As noted in Eq.~\ref{eq:spanning_number}, the average spanning number $n_{s}$ scales logarithmically with $L$ in the Goldstone phase and asymptotes to zero in the short loop phase.

\section{Measurement-induced phase transition in the Boundary}\label{sec:bulk_meas}

In this section, we investigate the occurrence of measurement-induced phase transitions in a 1D chain of qubits on the boundary of a toric code state, where the remaining qubits are measured in random local Pauli bases (as depicted in Fig. \ref{fig:adv}). Many of the findings in this section can be generalized to different topologies (such as torus, cylinder, or plane) and partitioning schemes of the unmeasured 1D chain.

We begin by examining a toric code implemented on a cylinder with two open circular boundaries. The boundary stabilizers are truncated, resulting in two types of boundary conditions: ``rough'' or ``smooth''. The rough condition arises when the plaquette stabilizers are truncated, while the smooth condition occurs when the star stabilizers are truncated. In Figure \ref{fig:surface_code}, the truncated toric code represented in the parton picture exhibits a simple form, where no distinction between rough and smooth is evident. Consequently, it is unnecessary to specify the type of boundary condition for the entanglement analysis. Moreover, various boundary conditions for the surface code can be created by measuring the toric code state on a torus, followed by applying a single layer of local unitary updates based on the measurement outcomes. For instance, to achieve a surface code with smooth boundary conditions, one should perform Pauli X measurements along a horizontal line on sub-lattice A. Conversely, a surface code with rough boundary conditions can be formed by conducting Pauli Z measurements along a horizontal line on sub-lattice B. Applying both of these measurement patterns results in the surface code illustrated in Fig. \ref{fig:surface_code}, featuring both rough and smooth boundary conditions. For concreteness, we will hereafter assume the pair of rough and smooth boundary conditions for the surface code.

\begin{figure*}[ht]
  \includegraphics[width=\textwidth]{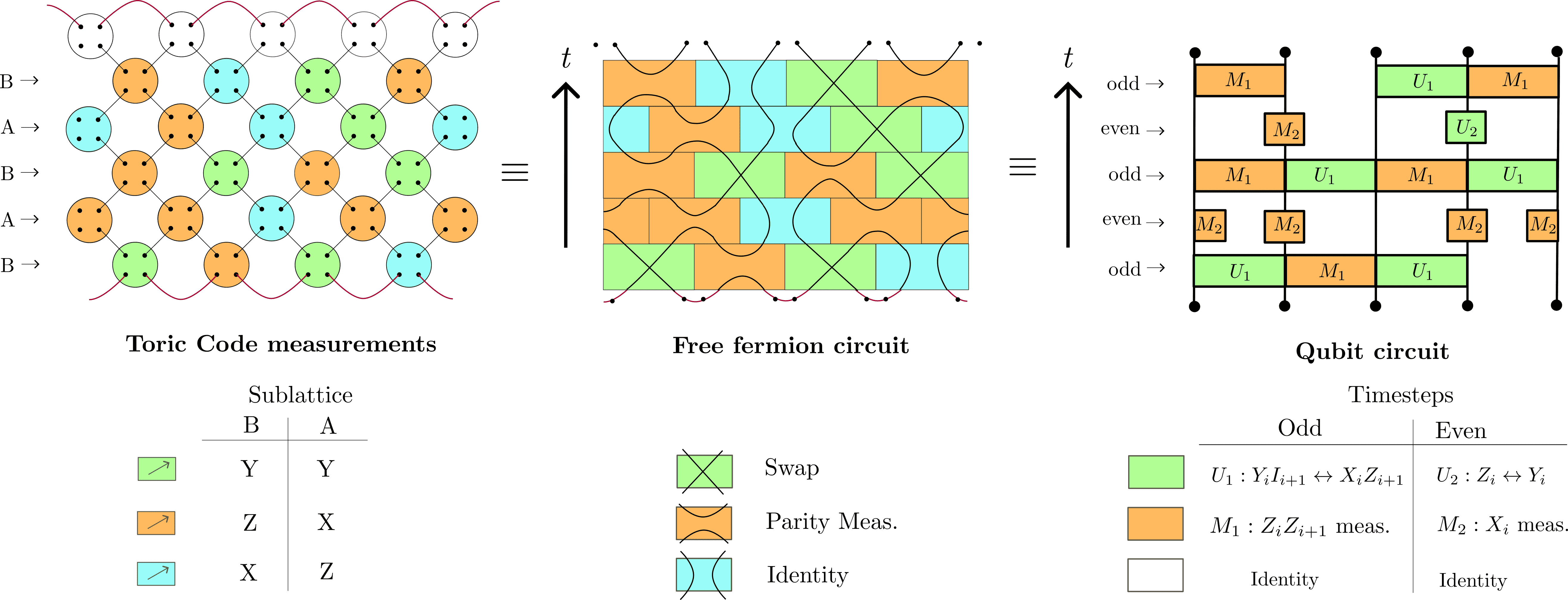}
  \caption{
  Pauli-$X/Y/Z$ measurements on the bulk, with the boundary chain of qubits left un-measured (left panel), correspond to a hybrid circuit with measurement and unitary gates. The measurement model is equivalent to a staggered measurement protocol on the Wen plaquette model. The plaquette model generates Majorana pairing patterns in that can be interpreted as world lines of Majorana fermions undergoing a free fermion hybrid circuit (middle panel). By Jordan-Wigner transformation, the same circuit can be identified with an Ising symmetric measurement and unitary circuit on qubits (right panel).}
  \label{fig:circuit_mapping}
\end{figure*}

In this section, we show that the boundary state after bulk measurements is precisely the state generated by a hybrid circuit in $1+1$d composed of measurements and unitaries. While this identification is general, for the particular case of the toric code, the corresponding hybrid circuit is a $1+1$d Ising symmetric circuit~\cite{Sang_2021, Sang_2020}. Our mapping is motivated by MBQC, whereby a circuit can be effectively realized by single-site measurements on a resource state. However, the toric code is not a universal resource state~\cite{Bravyi_2007}; thus, single-site measurement on the toric code cannot represent all circuits.

\subsection{Measured toric code as a $1+1$d hybrid circuit}\label{sec:bdy_circuit}


Starting from the parton representation of the toric code, measuring each qubit modifies the pairings of four neighboring Majoranas, as described in Sec.~\ref{sec:setup}.  These pairings can be interpreted as world lines of two Majoranas undergoing circuit operations. In particular, the corresponding circuit consists of the following gates acting on two neighboring Majorana fermions: swap gate, identity gate, and fermion parity measurement, as illustrated in the bottom of Fig.\ref{fig:circuit_mapping}.  Thus, starting from an $N\times N$ toric code state on a cylinder, the bulk measurements realize a depth $N$ free fermion circuit on $2N$ Majorana fermions (Fig.~\ref{fig:circuit_mapping} middle). Note that for a fixed configuration of measurement bases, different measurement outcomes correspond to hybrid circuits in $1+1$d differing only in the signs of the Majorana stabilizers. However, as noted earlier, the observables we are interested in--entanglement and spin-glass order--are not sensitive to these signs  and only depend on the worldline connectivity. Thus, for these observables, the hybrid circuits for a given measurement bases configuration are equivalent, regardless of the outcomes.




After all the projections onto the physical qubit Hilbert space are imposed, the Majorana hybrid circuit maps via Jordan-Wigner transformation to a depth $N$ hybrid circuit of local unitaries and local measurements on a one-dimensional system of $N$ qubits.  This mapping is shown explicitly in Fig.~\ref{fig:circuit_mapping}.  We note that this mapping is somewhat subtle, as the parton construction and Jordan-Wigner transformation are two different mappings from one qubit to respectively four and two Majorana fermions.  Briefly, the reason why the claimed mapping works is because at the top boundary, the top two of the four Majoranas per parton decomposition are always paired in a nearest neighbor dimer state (Fig.~\ref{fig:circuit_mapping} left), so the physical qubit state is solely determined by the bottom two Majoranas per site via the standard Jordan-Wigner mapping (see Appendix II~\cite{supp_mat} for details).       


The Majorana fermion parity measurements realize the measurements of either the neighboring $Z_{k}Z_{k+1}$ or on-site $X_k$ measurement in the qubit circuit, depending on which sub-lattice the measurements are performed:
\begin{eqnarray}
&M_{1} = i\tilde{\gamma}_{k,2}\tilde{\gamma}_{k+1,1} &= Z_{k}Z_{k+1} \nonumber\\
&M_{2} = i\tilde{\gamma}_{k,1}\tilde{\gamma}_{k,2} &= X_{k}. 
\end{eqnarray}
Similarly, the Majorana swap gate implements either a two-qubit unitary $U_{1}$ or on-site unitary $U_{2}$:
\begin{eqnarray}
&U_{1}: Y_{k}I_{k+1} \leftrightarrow X_{k}Z_{k+1} \nonumber\\
&U_{2}:Z_{k}\leftrightarrow Y_{k}. 
\end{eqnarray}

These circuit operations preserve an Ising $\mathbb{Z}_{2}$ symmetry 
$\prod_{i}X_{i}$.
The origin of this Ising symmetry of the hybrid circuit  
is the fact that the $\prod_{i}X_{i}$ string operator, supported on the rough boundary, takes definite value for the initial surface code state, and any bulk measurement away from the boundary commutes with this boundary operator. 

The qubit circuit corresponding to the $(p,q)$ measurement protocol on the toric code (TC) state can be explicitly defined as follows (see also Table~\ref{tab:circuit_mapping}):\\

(1) In odd times, perform one of the following operations on each pair of neighboring qubits: 2-qubit unitary $U_{1}$ (TC measurement along $Y$) with probability $p$, identity operation with probability $(1-p)q$ (TC measurement along $X$), and $M_{1}$ measurement with probability $(1-p)(1-q)$ (TC measurement along $Z$). \\

(2) In even times, perform one of the following operations on each qubit: on-site unitary $U_{2}$ with probability $p$(TC measurement along $Y$), on-site measurement $M_{2}$  with probability $(1-p)q$ (TC measurement along $X$), identity operation and  with probability $(1-p)(1-q)$ (TC measurement along $Z$).\\

\begin{table}
    \centering
    \begin{tabular}{c|c|c|c}
         Sublattice  & Measurement & Probability & Qubit Circuit  \\ (timesteps) &TC (WP) &  &  \\
         \hline
         \hline
         \multirow{3}{4em}{B (odd)}& $Y$ ($Y$) & $p$ & $U_{1}$\\         
         & $X$ ($Z$) & $(1-p)q$ & Identity\\
         & $Z$ ($X$) & $(1-p)(1-q)$ & $M_{1}$\\
         \hline 
         \hline
         \multirow{3}{4em}{A (even)}& $Y$ ($Y$) & $p$ & $U_{2}$\\
         & $X$ ($X$) & $(1-p)q$ & $M_{2}$\\
         & $Z$ ($Z$) & $(1-p)(1-q)$ & Identity\\
         \hline
    \end{tabular}
    \caption{Circuit mapping of the $(p,q)$ measurement model on the toric code (TC). For completeness, the corresponding staggered measurement protocol for the Wen plaquette (WP) model is mentioned in the parentheses.}
    \label{tab:circuit_mapping}
\end{table}



\subsection{Bipartite entanglement in the un-measured boundary}

By identifying the bulk Majorana pairings with the classical loop model as in Sec.~\ref{sec:loop_mapping}, we can establish a direct mapping between the entanglement across a bipartition of the boundary state and a specific quantity depicted schematically in Figure \ref{fig:phase_diagram}III within the loop model. This quantity, which counts the number of strands connecting two parts of the boundary chain, diagnoses the long-range correlations in the Goldstone phase of the loop model.

We also show in Appendix IA~\cite{supp_mat} that for the one boundary setup, there is a one-to-one correspondence between the Majorana strands and the canonical stabilizer generators after implementing projections. Specifically, a configuration with $n$ open strands connecting the two parts of the boundary corresponds to a quantum state with $n$ canonical stabilizer generators connecting the two parts, thereby contributing $\frac{n\ln{2}}{2}$ units of entanglement. 
The bipartite entanglement between two parts of the boundary thus acts as an order parameter for the phase transition in the loop configurations of the CPLC model.


In the Goldstone phase of CPLC,~\cite{Sang_2021} found that the entanglement between a contiguous sub-region $A$ of the qubit chain has a logarithmic scaling with a correction. Therefore, the entanglement of a contiguous subregion $A$ of the un-measured boundary chain of the toric code also satisfies the same entanglement scaling in the Goldstone phase,

\begin{equation}
\begin{split}
    \langle S_A \rangle \approx \frac{\ln(2)}{2}\left(\# \ \ln|A| + \frac{1}{4\pi}(\ln|A|)^2\right),
\end{split}
\end{equation}
while it obeys an area law in the short loop phase.

\section{Long-range order in the boundary state}\label{sec:order}



\subsection{Spin glass order parameter}

As we showed in the previous section, the bulk measurements performed on the toric code can be mapped to the Ising-symmetric hybrid circuits studied by Sang et al. \cite{Sang_2021, Sang_2020}.  Given only $ZZ$ measurements, the steady state is a ``random GHZ'' state characterized by a random spin configuration superposed with the flipped configuration.  This is also known as a spin glass state, and the spin glass order is captured by the 
Edwards-Anderson order parameter:
\begin{equation}\label{eq:ee_order}
O = \frac{1}{L}\sum_{i,j}^L\langle\psi|Z_iZ_j|\psi\rangle^2
\end{equation}
For spin glass order, $O\sim L$, whereas for paramagnetic order, $O\sim O(1)$.  

\cite{Sang_2020} studied the phase diagram of hybrid circuits involving $ZZ$ and $X$ measurements and random Ising-symmetric Clifford unitaries, and a stable spin glass phase was found.  The toric code measurements map to a subset of symmetric Clifford unitaries-- namely the free fermion operations defined above-- and for this restricted class we provide a fermionic perspective on the spin glass order parameter and the extent of the spin glass phase.

The central object in the spin glass order parameter is $Z_iZ_j$, which for a stabilizer state can take three values ($\pm 1$ or $0$).  It maps via Jordan-Wigner transformation to 
\begin{equation}\label{eq:zizj}
Z_iZ_j = i\gamma_{i,2} \left(\prod_{k=i+1}^{k=j-1}i\gamma_{k,1}\gamma_{k,2}\right) \gamma_{j,1}.
\end{equation}
This string of Majoranas is nonzero if and only if  
all Majoranas within the interval $(i,j)$ are paired amongst themselves.  (If any Majorana within the interval is paired with one outside, that dimer will anti-commute with the above string and render its expectation value zero.) 

In the short loop phase, configurations are composed of loops with a characteristic size $\xi$, which is independent of $L$. Hence, the probability that all Majoranas within an interval $(i,j)$ are paired up is independent of $|i-j|$ for $|i-j|\gg \xi$.  For $q<1/2$, this probability is a nonzero $O(1)$ number independent of $|i-j|$, and thus the spin-glass order parameter is extensive in the $q<1/2$ short loop phase.  


\subsection{Linear order parameter from adaptive circuits}

Due to the equal probabilities of $Z_iZ_j$ having opposite signs $\pm 1$, its average value is zero, making it necessary to use nonlinear order parameters such as the Edwards-Anderson order parameter defined in Eq.~\ref{eq:ee_order}. However, in experimental setups, measuring nonlinear order parameters is generally challenging and requires post-selection of the measurement outcomes.  In practice it is more feasible to access expectation values of operators like $\mathrm{Tr}(\rho O)$ which are linear in the density matrix $\rho$, the ensemble of all measurement trajectories. 


Here we detail an efficient protocol for converting the (nonlinear) spin glass order into a linear order parameter. Our protocol employs an adaptive strategy that applies a layer of local unitaries on the boundary conditioned on the bulk measurement outcomes. The main objective of this adaptive unitary layer is to transform all the $\pm$ values of $Z_{i}Z_{j}$ stabilizer generators into positive values, thus eliminating the issue of sign cancellation and converting the spin glass order into long-range ferromagnetic order, which is linear in the state and can be accessed experimentally. The protocol consists of two main parts: (1) identify $Z_iZ_j$ stabilizers and determine their signs, and (2) obtain single-site unitaries that can correct the negative signs to positive signs.

Note that one approach for identifying and correcting the sign of  $Z_{i}Z_{j}$ operators in the stabilizer group is to simulate the entire evolution classically using the stabilizer formalism, with the knowledge of all the $O(L^{2})$ bulk measurement outcomes. However, we propose a simpler algorithm that only requires access to the directions and outcomes of measurements within a correlation length $O(\xi)$ from the boundary, i.e. $O(L\xi)$ measurements. 


\begin{figure}
    \centering    \includegraphics[width = 0.9\columnwidth]{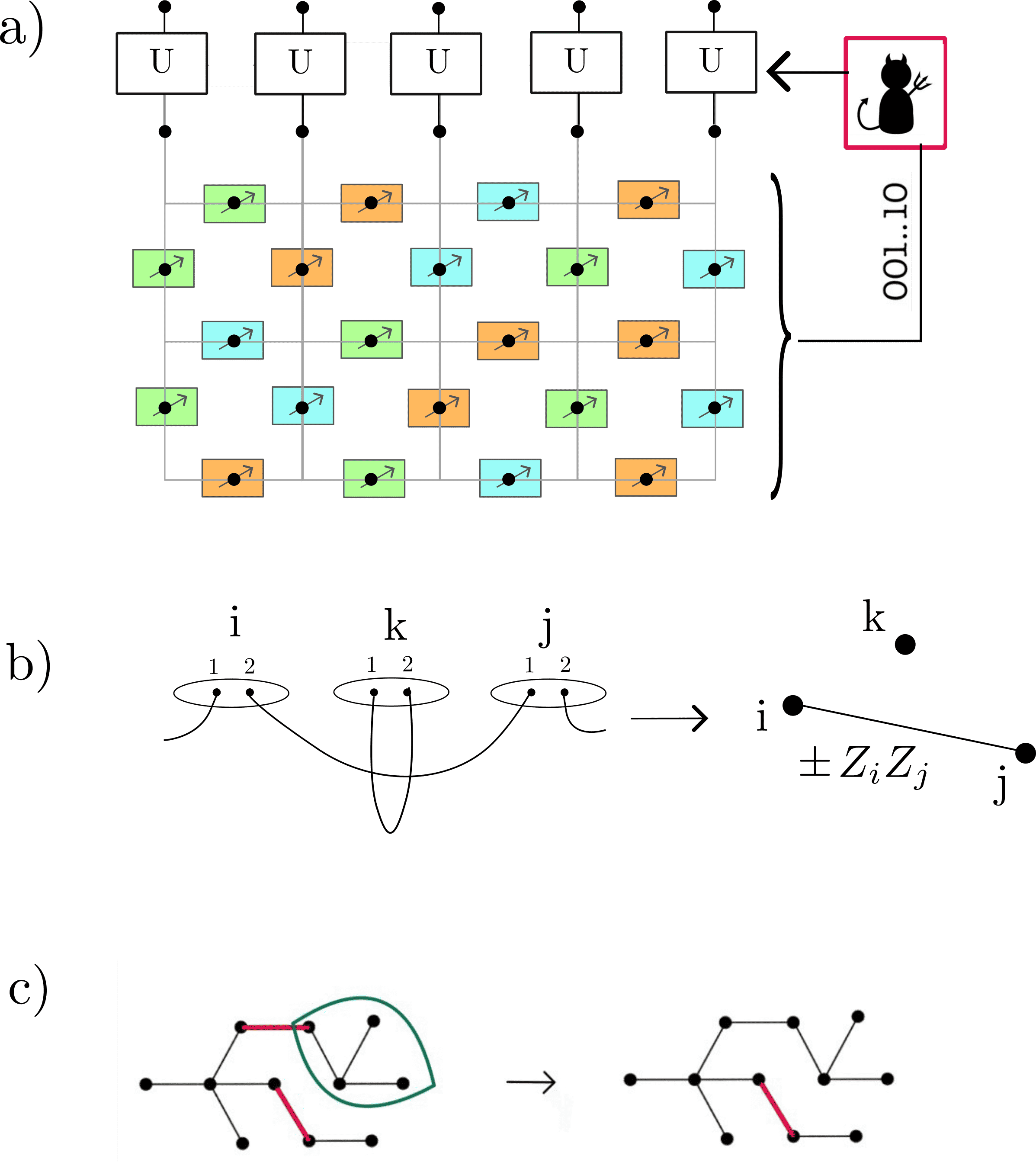}
    \caption{(a) To convert any (nonlinear) spin glass long-range order of the boundary state into (linear) ferromagnetic long-range order, a layer of on-site unitaries conditioned on bulk measurement outcomes can be applied to the boundary. (b) The classical processing for the adaptive protocol involves using the distribution of Majorana strands to construct a graph in which an edge between nodes (qubits) $i,j$ represents existence of a $Z_iZ_j$ stabilizer. The edges are colored depending on whether the sign of the stabilizer is $\pm 1$, which can be computed from the measurement outcomes along the strands. (c) The negative signs in the tree graph can be flipped by applying $X_{k}$ on all nodes on one side of the negative edge (e.g. those encircled).}
    \label{fig:demon}
\end{figure}


The algorithm is as follows:

\textbf{A. $Z_{i}Z_{j}$ generator graph construction:}
We construct a graph whose vertices are the boundary qubit sites $i$ and which has an edge $ij$ if $\pm Z_{i}Z_{j}$ is in the stabilizer group.  By knowing the positions of $X,Y,Z$ measurements, we can obtain the corresponding configuration of Majorana strands.  If for any $i<j$, $\gamma_{i,2}$ is paired up with $\gamma_{j,1}$, we draw an edge $ij$ in the graph. As per Eq.~\ref{eq:zizj}, this pairing implies a $Z_{i}Z_{j}$ stabilizer if and only if all the Majorana fermions in the interval $(i,j)$ are paired up internally. This can be checked for all the Majorana fermions in $(i,j)$ from the loop configuration. If indeed there are strands that exit the interval $(i,j)$, then we erase the edge $ij$ in the graph as this doesn't correspond to a $Z_{i}Z_{j}$ stabilizer. For the special case $p = 0$ without loop crossings, this second step is not necessary, as all the Majorana strands must be nested in this case. Note that the graph is a tree as the Majorana strands are all independent generators. 

We then obtain the sign of the $Z_{i}Z_{j}$ stabilizers by tracking the sign of the arrows along the strand $\gamma_{i,2}\gamma_{j,1}$ and all the intermediate strands $\gamma_{k,1}\gamma_{k^{\prime},2}$ for $k,k^{\prime}\in (i,j)$. This requires us to keep track of only $O(L\xi)$ measurement outcomes, where $\xi$ is the correlation length corresponding to the loop size in the underlying CPLC model.  If $Z_{i}Z_{j}=-1$, we color the edge $ij$. 

\textbf{B. Correcting the sign:} From the previous step, we have a colored graph. Flipping a spin $k$ (acting with the unitary $X_{k}$) changes the signs of all adjacent edges to the node $k$. We can correct any negative sign on an edge by flipping all nodes on one side of the edge (see Fig. \ref{fig:demon}c). Repeating this process for each edge allows us to flip the sign of every edge individually. To correct all the edges with a minimal number of operations, a search algorithm within the tree can be performed, which has a polynomial complexity with respect to the system size. 

Given access to the measurement protocol and outcomes, this algorithm (with complexity polynomial in system size) can be executed by a classical computer to determine the necessary adaptive unitary protocol, resulting in all trajectories having non-negative $Z_iZ_j$ correlations.  The resulting ensemble of trajectories $\rho$ has long-range ferromagnetic order $\langle Z_i Z_j\rangle_\rho$ in the original spin glass phase.

\section{General on-site measurements}\label{sec:beyond_cplc}

We now consider the effect of measurements along an arbitrary direction $n_{x}X+n_{y}Y+n_{z}Z$ on the toric code. These map to statistical models that go beyond the scope of the previously discussed CPLC model (Section~\ref{sec:loop_mapping}). In the following, we will demonstrate the representation of the measured toric code using Gaussian tensor networks (GTN) and the realization of Gaussian hybrid circuits within the virtual space. Furthermore, we will establish a connection with well-studied Gaussian hybrid circuits to showcase the robustness of the boundary MIE phase diagram obtained from the CPLC model in the general measurement case.

\subsection{Tensor network representation of parton construction}

The parton representation of the toric code state can be reinterpreted as a two-dimensional tensor network composed of local tensors $\ket{T}$. These tensors consist of a single physical leg representing a qubit and four virtual legs representing the parton Majorana fermions. 
To construct a tensor network, contractions between spins or contractions between Majorana fermions are allowed.  A contraction between spins involves projecting the two legs onto a maximally entangled state, and a contraction between Majorana fermions sharing an edge involves projecting the two Majorana fermions onto the eigenstate of the $i\gamma_i\gamma_j$ operator with eigenvalue +1.  The latter assumes an orientation for each virtual bond which must be specified. 



\begin{figure}
    \centering
    \includegraphics[width = \columnwidth]{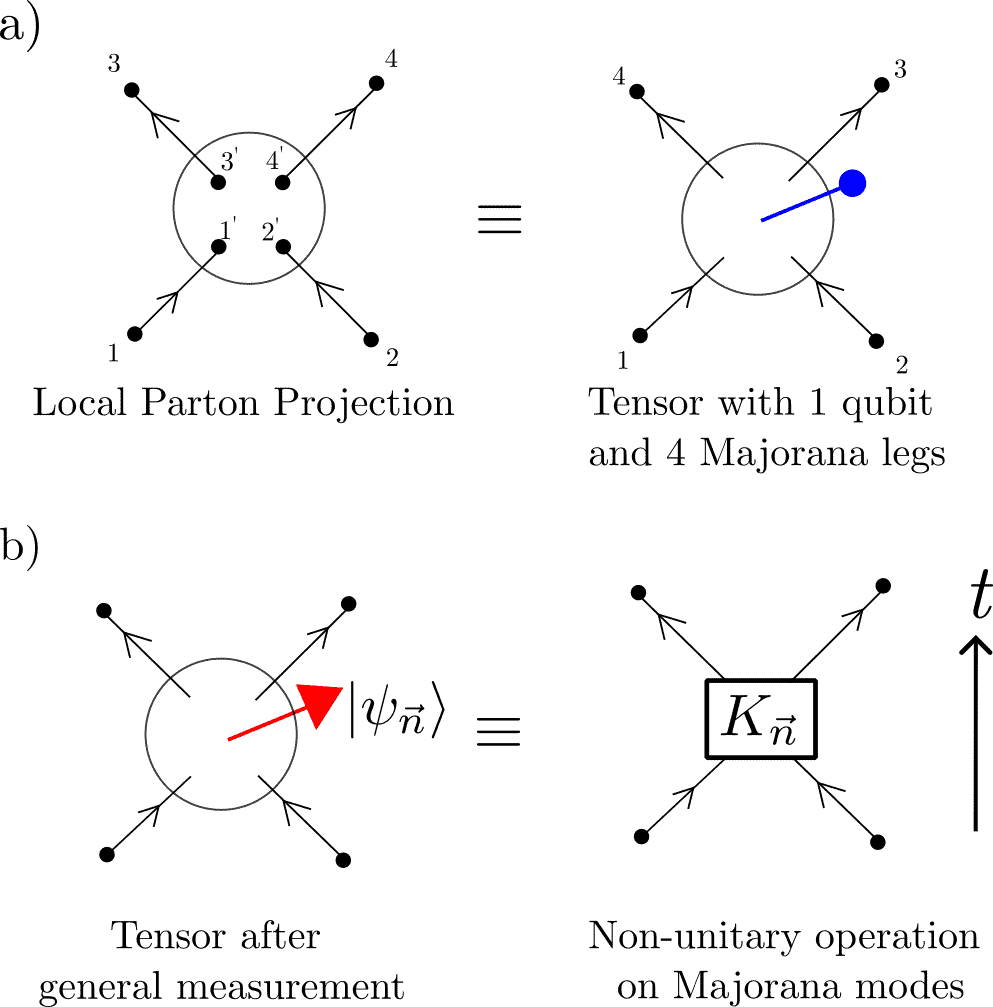}
    \caption{(a) The parton representation of the toric code state can be interpreted as a local tensor with 1 `physical' qubit leg and 4 `virtual' Majorana legs, along with an orientation for contraction of the virtual indices. In this case, the orientation corresponding to a toric code groundstate is specified. (b) Measuring the parton state, or equivalently, contracting the physical index with a state vector in a general direction, leads to a Gaussian state on the 4 `virtual' Majorana degrees of freedom. This Gaussian state can be mapped to a Gaussian operation on 2 Majorana fermions, and can be directly identified with a non-unitary operator.}
    \label{fig:gaussian_tn}
\end{figure}

We now describe the tensor network representation of the projected parton states of the toric code. First, we introduce the projection tensor that maps four Majorana fermions to one spin:
\begin{equation}
P = \frac{\ket{G_1} \ket{\uparrow} + \ket{G_2} \ket{\downarrow}}{\sqrt{2}}. 
\end{equation}
Here $\ket{G_1}$ and $\ket{G_2}$ are two fermionic stabilizer states of the four Majoranas, where $\ket{G_1}$ is stabilized by $i\gamma_{1}\gamma_{3}$, $i\gamma_{2}\gamma_{4}$ and $\ket{G_2} = i\gamma_{1}\gamma_{2}\ket{G_1}$. The orientation for bonds between tensors is specified in Fig \ref{fig:gaussian_tn} (a), 
and corresponds to the orientation of Majorana pairs  described in Section \ref{sec:setup}.


\subsection{Measured toric code as a Gaussian tensor network}

Measuring a qubit in the $\Vec{n}$ axis and obtaining outcome $\pm1$ corresponds to contracting the physical leg of a tensor with the qubit state $\ket{\psi_{\pm\Vec{n}}}$, the eigenstate of operator $\Vec{n}\cdot \Vec{\sigma} = n_{x}X+n_{y}Y+n_{z}Z$ with eigenvalue $\pm1$.  After contraction, the resulting projection tensor takes the form of a Gaussian fermionic tensor on the remaining Majorana legs:
\begin{equation}
\begin{split}
\ket{F_{\Vec{n}}} = \frac{\ket{G_1} \bra{\psi_{\Vec{n}}}\ket{\uparrow} + \ket{G_2} \bra{\psi_{\Vec{n}}}\ket{\downarrow}}{\sqrt{2}} \\
=\frac{e^{i\phi}\cos{\frac{\theta}{2}}\ket{G_1} + \sin{\frac{\theta}{2}}\ket{G_2}}{\sqrt{2}}
\end{split}
\end{equation}

Here, $\theta$ and $\phi$ represent the spherical coordinates of the unit vector $\Vec{n} = (n_{x},n_{y},n_{z})=(\sin\theta\cos\phi,\sin\theta\sin\phi,\cos\theta)$. The fermionic operators $F^1$ and $F^2$ which stabilize $\ket{F_{\Vec{n}}}$ (meaning that $F^{1,2}\ket{F_{\Vec{n}}}=\ket{F_{\Vec{n}}}$) are:
\begin{equation}
\begin{split}
F^1 = i\gamma_1\gamma_3 \cos{\theta}+i\gamma_1\gamma_2\sin{\theta}\cos{\phi}+i\gamma_1\gamma_4\sin{\theta}\sin{\phi}\\
F^2 = i\gamma_2\gamma_4 \cos{\theta}+i\gamma_4\gamma_3\sin{\theta}\cos{\phi}+i\gamma_3\gamma_2\sin{\theta}\sin{\phi}.\nonumber
\end{split}
\end{equation}
The independence, commutation, and unit-square properties of the fermionic operators $F^1$ and $F^2$ can be straightforwardly demonstrated. These two fermionic operators uniquely define $\ket{F_{\Vec{n}}}$ as a Gaussian state/tensor which in fact is the most general form for four Majorana fermions. The covariance matrix of the state (which is defined as $\Gamma_{ij} = \langle \frac{i}{2}[\gamma_{i},\gamma_{j}]\rangle$) in the $(\gamma_{1},\gamma_{2},\gamma_{3},\gamma_{4})$ basis is:

\begin{equation}
    \Gamma = \frac{1}{2}\begin{bmatrix}
0 & \sin{\theta}\cos{\phi} & \cos{\theta} & \sin{\theta}\sin{\phi} \\
-\sin{\theta}\cos{\phi} & 0 & -\sin{\theta}\sin{\phi} & \cos{\theta} \\
-\cos{\theta} & \sin{\theta}\sin{\phi} & 0 & -\sin{\theta}\cos{\phi} \\
-\sin{\theta}\sin{\phi} & -\cos{\theta} & \sin{\theta}\cos{\phi} & 0 \\
\end{bmatrix}\nonumber
\end{equation}

In summary, each qubit measurement in the $\Vec{n}$ direction results in a Gaussian tensor supported on virtual legs. In the setup where all but a top boundary of qubits are measured, all degrees of freedom except the top boundary of qubits are contracted (Fig. \ref{fig:gaussian_ckt}).  We show in appendix \cite{supp_mat} that the Jordan-Wigner transformation of this boundary qubit state precisely yields the Gaussian tensor network state in which the boundary qubit projections and top row of Majorana contractions are removed (Fig.\ref{fig:gaussian_ckt} right).  Consequently, performing measurements in the bulk of the toric code and employing a Jordan-Wigner transformation at the boundary results in a  remaining state described by a Gaussian tensor network. 

In the measured toric code, each quantum trajectory can be represented by ${\vec{n}_i}$, where $\vec{n}_i$ indicates both the type and the outcome of the measurement when measuring the qubit at site $i$. In this notation, the type of measurement is encoded in the axis of ${\vec{n}}$, denoted as $e_{\vec{n}}$, while the measurement result is encoded in the direction of ${\vec{n}}$ along the axis. The probability of each trajectory arises from two sources. Firstly, there is the classical probability set by the protocol, denoted as $w\left(e_{\vec{n}}\right)$, representing the probability of selecting measurement axis $e_{\vec{n}}$. If the measurements bases are chosen independently from site to site, the probability of a given set of measurement bases is $w\left(\{e_{\vec{n}_{i}}\}\right) = \prod_i w\left(e_{\vec{n}_{i}}\right)$.  Secondly, the Born rule for the measurement outcome determines the probability $P(\vec{n})$ of the sign associated with $\vec{n}$ based on the measurement. The Born probability $P(\{\vec{n}_i\})$ of each trajectory $\{\vec{n}_i\}$ is the norm of the wave function represented by the tensor network.
 The combined probability for a particular quantum trajectory is thus $w(\{e_{\vec{n}_{i}}\})P(\{\vec{n}_{i}\})$.

This notation can be used to represent the $(p,q)$ measurement protocol for $Z$, $Y$, or $X$ measurements discussed in previous sections. Due to axis based definition of $w$ and the fact that $e_{\vec{n}} = e_{-\vec{n}}$, the probability distribution trivially satisfies $w\left(e_{\vec{n}}\right) = w\left(e_{-\vec{n}}\right)$. Therefore, the weights associated with these measurements are $w\left(e_Z\right) = w\left(e_{-Z}\right)= (1-q)(1-p)$, $w\left(e_Y\right) = w\left(e_{-Y}\right) = p$, $w\left(e_X\right)= w\left(e_{-X}\right) = q(1-p)$. Based on the stabilizer formalism, the Born probability for each plus and minus sign is equal to $P(\vec{\sigma}) = \frac{1}{2}$. Consequently, $\sum_{\vec{\sigma}}w\left(e_{\vec{\sigma}}\right)P(\vec{\sigma}) = 1$. Also note that the classical probability satsifies: $\sum_{e_{\vec{\sigma}}}w\left(e_{\vec{\sigma}}\right)= w\left(e_{X}\right) + w\left(e_{Y}\right) + w\left(e_{Z}\right) = 1$.
\subsection{Measured toric code as a Gaussian hybrid circuit}

\begin{figure}
    \centering
    \includegraphics[width = \columnwidth]{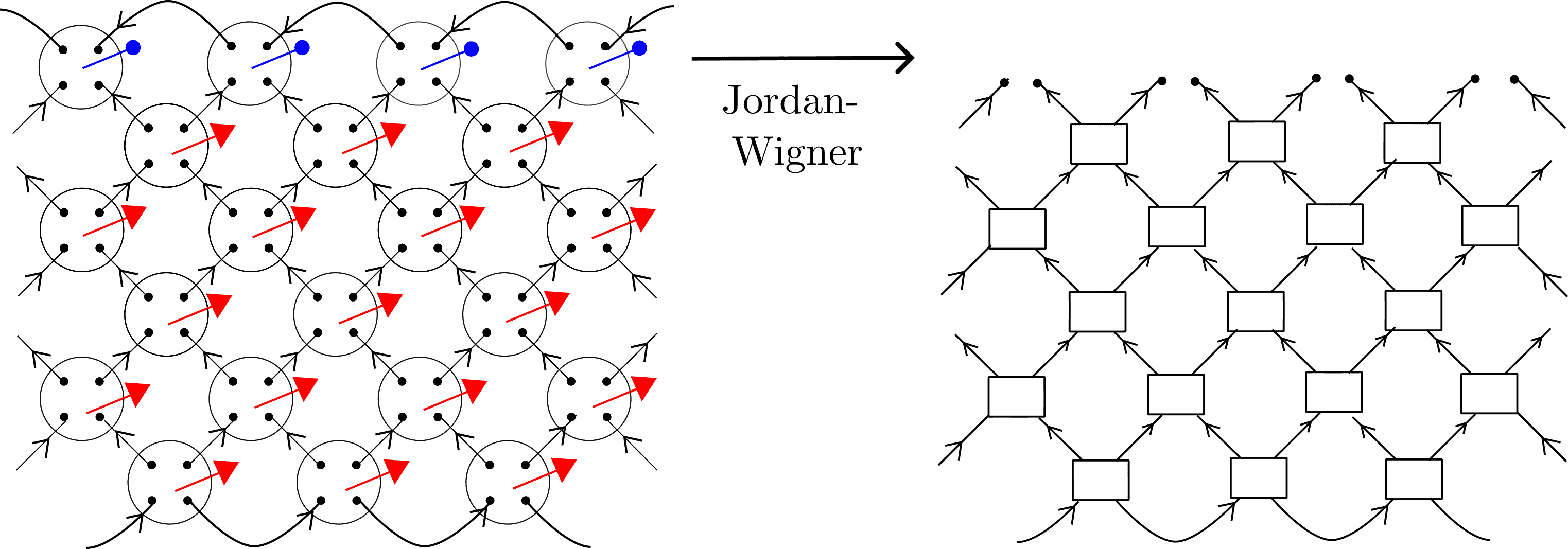}
    \caption{Bulk measurements of general on-site operators on the toric code ground-state can be mapped to a Gaussian circuit on a chain of the virtual Majorana degrees of freedom. The qubit state supported on the top boundary on the left hand side maps under Jordan-Wigner transformation to the Gaussian state supported on the top boundary on the right hand side (see Appendix III~\cite{supp_mat} for derivation).}
    \label{fig:gaussian_ckt}
\end{figure}


The Gaussian tensors in the virtual space can be understood as Gaussian operations that perform non-unitary transformations between the lower and upper legs of the virtual space, as illustrated in Figure \ref{fig:gaussian_tn}(b). Building upon this observation, we establish a correspondence between a specific class of non-unitary Gaussian circuits in 1+1D and the random Gaussian tensor networks in 2D discussed in the previous section. 

These Gaussian circuits are characterized by a set of operators $\{K_{\Vec{n}}\}$ and a probability distribution $\tilde{w}(\Vec{n})$ associated with the operations. The operators $K_{\Vec{n}}$ act on the neighboring Majorana modes $\gamma_{i}$ and $\gamma_{i+1}$ within the virtual space, resulting in non-unitary Gaussian circuits~\cite{jian2022criticality,jian2023measurementinduced}. More explicitly, these operators can be expressed as:

\begin{equation}
K_{\Vec{n}} = \left(1-n_{x}^{2}\right)^{1/4}e^{-i\alpha(\Vec{n})\gamma_{i}\gamma_{i+1}}
\end{equation}
where $\alpha(\Vec{n})$ is defined by 
\begin{equation}
e^{-2\mathrm{Re}[ \alpha(\Vec{n})] } = \left(\frac{1+n_{x}}{1-n_{x}}\right)^{1/2};
e^{i2\mathrm{Im} [\alpha(\Vec{n})]} = \frac{in_{y}+n_{z}}{\left(n_{y}^{2}+n_{z}^{2}\right)^{1/2}}
\label{eq:gaussian_ckt_mapping}
\end{equation}

It is worth noting that when $n_x$ is zero, the operations are unitary. However, when $n_x$ is $\pm1$, the operations correspond to fermion parity measurements. For general $\vec{n}$ with $n_{x} \neq 0$, the non-unitary operation $K_{\vec{n}}$ is a weak measurement of the fermion parity. The spacetime geometry of the monitored Gaussian circuit acting on a Majorana chain is depicted in the right hand-side of Fig \ref{fig:gaussian_ckt}, and each operation is randomly chosen from the set of operations $K_{\Vec{n}}$. The probability of each trajectory is determined by two sources: the classical distribution $\tilde{w}(\vec{n}_{i})$ and the Born probability $ \bra{\psi}K_{\vec{n}_{i}}^{\dagger}K_{\vec{n}_{i}}\ket{\psi}$, where $\ket{\psi}$ is the normalized wave function of the 1D fermion chain before the action of the operator $K_{\vec{n}_{i}}$, and the updated wave function after the action is given by $\frac{K_{\vec{n}_{i}}\ket{\psi}}{||K_{\vec{n}_{i}}\ket{\psi}||}$. Note that the classical probability should satisfy $\tilde{w}((n_x,n_y,n_z)) = \tilde{w}((-n_x,n_y,n_z))$ to not force any bias on the measurement outcomes.

The first step in establishing the correspondence is to relate the operator representation of the tensor $\ket{F_{\Vec{n}}}$ in the tensor network to the set of operators in the Gaussian circuit. By performing explicit contractions of the tensor network, it can be shown that the operator representation of the tensor $\ket{F_{\Vec{n}}}$ is $\frac{K_{\vec{n}}}{\sqrt{2}}$. This implies that the final state of the hybrid circuit and the tensor network, given the same set of directions, are equal up to a normalization factor. However, this normalization factor is only relevant for the probability of the trajectory. This mapping establishes a direct correspondence between the measurement directions and outcomes (labeled by $\Vec{n}$) and a family of operators in the circuit representation parameterized by $\Vec{n}$ (Fig.\ref{fig:gaussian_ckt}). 

The second step is to establish a connection between $\tilde{w}(\Vec{n})$ and $w(e_{\Vec{n}})$ such that identical space-time configurations in both setups have the same probability. The probability of a specific trajectory in the tensor network can be expressed as $P_{\{\vec{n}_{i}\}} = \prod_i w\left(e_{\vec{n}_{i}}\right)\bigg|\bigg|\prod_{i}\frac{K_{\vec{n}_{i}}}{\sqrt{2}}\ket{\psi_0}\bigg|\bigg|^2$, $\ket{\psi_0}$ being the initial state / lower boundary of the toric code. By recursively defining $\ket{\psi_t} = \frac{K_{\vec{n}_{t}}\ket{\psi_{t-1}}}{||K_{\vec{n}_{t}}\ket{\psi_{t-1}}||}$ the probability can be written as: $P_{\{\vec{n}_{i}\}} = \prod_t \frac{w\left(e_{\vec{n}_{i}}\right)}{2}||K_{\vec{n}_{t}}\ket{\psi_{t-1}}||^2$, which establishes the correspondence with the trajectory probability in the circuit representation.

The $w(e_{\Vec{n}})$ ensemble of measurements on toric code is thus the same as the $\tilde{w}(\Vec{n}) = \frac{w(e_{\Vec{n}})}{2} $ ensemble of non-unitary circuits on Majorana fermions, where the Gaussian operations $K_{\vec{n}}$ are sampled with protocol probability $\tilde{w}(\vec{n})$ (with the restriction $\tilde{w}\left(\vec{n}\right) = \tilde{w}\left(-\vec{n}\right)$). This establishes the mapping between the measurement protocol on toric code and a specific class of non-unitary Gaussian circuits on the virtual space. However, it is important to note that the integration domains for $\tilde{w}(\Vec{n})$ and $w(e_{\Vec{n}})$ are different. While $w(e_{\Vec{n}})$ is defined over a hemisphere due to its axis-based definition, $\tilde{w}(\Vec{n})$ is defined over the entire sphere.

\subsection{Phase diagram of boundary MIE after general on-site measurement}
Following the circuit mapping that connects different measurement protocols on toric code with Gaussian fermionic circuits, we can readily use results of the entanglement phase diagram of non-unitary Gaussian circuits to infer the phase diagram of the MIE in the boundary state after measuring the bulk of the toric code state on a cylinder, along any general directions. The corresponding circuit problem has been extensively studied both numerically and analytically recently, for e.g. see ~\cite{Merritt_2023,jian2022criticality,jian2023measurementinduced}. Corresponding to any Gaussian circuit ensemble that satisfies the condition $\tilde{w}\left(\vec{n}\right) = \tilde{w}\left(-\vec{n}\right)$, we can find the bulk measurement protocol on the toric code that realizes that Gaussian circuit, via Eq.~\ref{eq:gaussian_ckt_mapping}. 

These works have found the entanglement phase diagram to consist of regions of area law and critical logarithmic scaling separated by phase transitions, and we infer that the boundary MIE phase diagram also behaves similarly. This demonstrates that the MIE phase diagram we obtained by mapping the result of specific measurement protocol (along $X,Y, $ or $Z$ directions) to the CPLC model, is qualitatively robust to modifications of the measurement protocol to general on-site measurements. However, the specific phase boundaries and the nature of the phase transition between the area law and critical phases vary between different ensembles, as discussed in~\cite{jian2023measurementinduced}.

\section{Discussion}\label{sec:discussion}
We explored the use of the measurement-based quantum computing (MBQC) setup for generating and manipulating quantum phases of matter. Specifically, we focused on the toric code, a topologically ordered state, and mapped the effects of random Pauli measurements to a classical loop model, allowing for an analytical understanding for measurement-induced entanglement. Additionally, we mapped general on-site measurements to Gaussian tensor networks and hybrid circuits.  

We found that the entanglement pattern imprinted on the un-measured qubits following measurement of the bulk of the toric code groundstate undergoes a phase transition that reflects the transition in the corresponding classical loop model. When a boundary chain of qubits is left un-measured, the boundary state can have either area law or logarithmic scaling of entanglement entropy, depending on the relative $X,Y,Z$ Pauli measurement frequencies. Additionally, we found that these states can also be distinguished by a spin-glass order parameter. This allowed us to devise an adaptive protocol conditioned on the measurement results, which can steer the boundary state into a ferromagnetically ordered state, and this can be efficiently probed in experiments. 

Because of the relative simplicity of bulk single-site measurements and the fact that toric code states have already been realized in quantum hardware \cite{toric1, toric2}, the setup described in this work  are experimentally relevant. 
Our MBQC-based setup provides a way to simulate $d+1$-spacetime dimensional hybrid circuits by one layer of local measurements on a $d+1$ dimensional entangled resource states. In quantum devices with limited coherence times, this setup may provide a promising practical route towards simulating such hybrid circuits. 



Our work illustrates how parton constructions can be leveraged in MBQC schemes, and it is worth exploring generalizations especially in higher-dimensional resource states. For example, fracton orders admit Majorana parton descriptions \cite{fracton1, fracton2}, and one can consider the effect of measurements on such states.  Another noteworthy example is the Levin-Wen 3D plaquette model \cite{Levin_2003}, which serves as a natural generalization of the projection of a free parton state prepared on the edges. One can analyze the effect of measurements by a very similar mapping to a loop model in three dimensions. Furthermore, considering higher dimensional un-measured manifolds might lead to more complex entanglement structures. 

It is also interesting to investigate the relation between measurement-induced entanglement (MIE) in random bases to other aspects of wavefunction complexity.  For example, MIE after measurements in a fixed basis can diagnose the sign structure in that particular basis \cite{Lin_2023}, and randomizing the measurement bases may partially probe the robustness of the sign structure to local unitary transformations (``intrinsic sign structure''). There may also be connections between the universality of the resource state in MBQC and the entanglement pattern induced by measurement. For example, a similar bulk measurement protocol on cluster states, which are universal MBQC resources, leads to a phase transition between area and volume-law states~\cite{Liu2022}. This can be contrasted with the toric code case (which is not a resource for universal MBQC) in this work, where any subregion of the boundary state has at most logarithmic scaling of entanglement entropy.


\subsection{Acknowledgments}
We thank Xiao Chen, Paul Herringer, Peter Lu, Adam Nahum, and Beni Yoshida for useful discussions and Amin Moharramipour for carefully reading our manuscript.  This work was supported by the Perimeter Institute for
    Theoretical Physics (PI), the Natural Sciences and Engineering Research Council of Canada (NSERC), and an Ontario Early Researcher Award. Research at PI is supported in part by the Government of Canada through the Department of Innovation, Science and Economic Development
    and by the Province of Ontario through the Ministry of Colleges and Universities.

\nocite{*}

\providecommand{\noopsort}[1]{}\providecommand{\singleletter}[1]{#1}%

\newpage
\onecolumngrid
\appendix

\section{Loop patterns and quantum states in physical Hilbert space}\label{appsec:projection}

Here we establish a correspondence between a loop pattern in the system and the Majorana stabilizer generators within the physical Hilbert space. We emphasize that the loop patterns arise as a consequence of measuring the free Majorana state outside of the physical Hilbert space. However, it is crucial to note that even after these measurements, the resulting states must be projected back into the physical Hilbert space. By utilizing a loop pattern configuration, we demonstrate the construction of stabilizer generators in a canonical form, which allows us to quantify bipartite entanglement by simply counting the number of stabilizer generators connecting the two parts.

\begin{figure}[h]
    \centering    \includegraphics[width = 0.7\columnwidth]{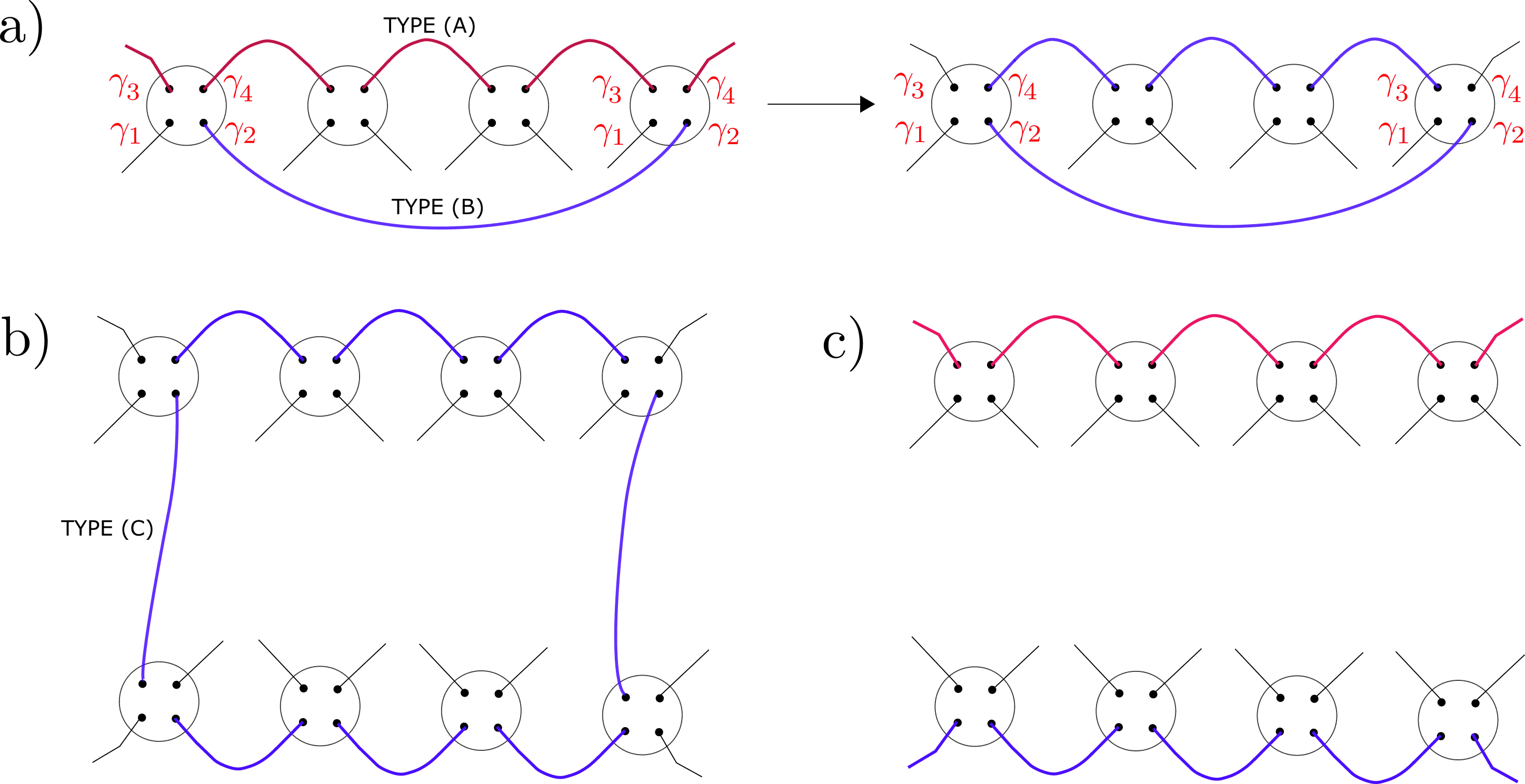}
    \caption{\textbf{One un-measured boundary:}(a) Stabilizer generators of the boundary state from a given pattern of loops can be obtained by multiplying a type (B) pairing (blue strand on the left) with all the interior type (A) strands (colored red in the left). \textbf{Two un-measured boundaries:} (b) Stabilizer generator obtained by multiplying two (C) type strands with all intervening (A) strands in upper and lower boundaries. (c) Two independent stabilizer generators which are products of the type (A) strands in upper and lower boundaries. }
    \label{fig:bdy_op_update}
\end{figure}

\subsection{Single un-measured boundary}\label{appsec:projections_1bdy}
Consider first the scenario of measuring all the bulk qubits of the Toric code, while leaving just a chain of $N$ un-measured spins on the boundary. Using the parton picture, we find that the Majorana fermions on the boundary are paired up using two types of Majorana strands: (A) which connect the nearest neighbor $\gamma_{3}$ and $\gamma_{4}$ Majoranas, and (B) which connect the $\gamma_{1}$ and $\gamma_{2}$ operators through the bulk (see Fig.~\ref{fig:bdy_op_update}a).

Note, however, the $N$ type (A) operators don't belong to the physical Hilbert space (set by $D_{j} = 1$), as they anti-commute with the Projection operators (although, the product of the type (A) operators belong to the stabilizer group as they commute with projection operators, assuming periodic boundary condition.). However, for each type (B) pairing, we can construct a stabilizer generator for the state by multiplying all the type (A) pairings in its interior, which now commutes with all the on-site projection operators. Since there must be $N$ type (B) pairings, we can form $N$ stabilizer generators, which completely define the state on the boundary. Furthermore, these stabilizer generators are already in the canonical form defined in the main text, i.e. if we take the subset of stabilizer generators with support on any interval and truncate them to that interval, they form an independent set. This implies that if in a particular loop configuration there are $n$ pairings between two parts of the system $A$ and $\overline{A}$, the corresponding quantum state has an entanglement between $A$ and $\overline{A}$ which is $\propto n$.

\subsection{Two un-measured boundaries}\label{appsec:projections_2bdy}
This section focuses on the case where both boundaries of the cylinder are un-measured, and examines the impact of physical Hilbert space projection on the stabilizer generators. Let us call the two boundaries `upper' and `lower'.  We are interested in calculating the entanglement between the upper and lower boundaries.

As in the previous subsection, there will be the types (A) and (B) pairings on both the boundaries (see Fig \ref{fig:bdy_op_update}b). Again, as before, we can remove the (A) pairings, and form stabilizer generators for each of the (B) type pairings. Note however, none of the (B) type pairings have support on both upper and lower boundaries, and thus don't contribute to any entanglement between them.

In this scenario, there is yet another type of pairing, which pair up $\gamma_{1,2}$ Majoranas between the upper and lower boundaries, which we call type (C). Note, by construction, there can only be even number of such strands. These pairings do not represent physical operators by themselves as they anti-commute with the projection operators. However, we can form physical operators by multiplying two type (C) strands with all the intervening type (A) strands in the upper and lower boundaries. This operator is a physical stabilizer generator, since it intersects all projection operators an even number of times (see Fig.~\ref{fig:bdy_op_update}b).

Suppose there are $n$ type (C) pairings in a configuration. If $n = 0$ the entanglement between upper and lower boundaries is clearly zero, as no stabilizer generator in the canonical form has support on both the boundaries. For $n\geq 2$, we can form $n$ such stabilizer generators by the above recipe, as in Fig.~\ref{fig:bdy_op_update}b. Note, however, all these generators are not independent. In particular, there are two generators with support only on either upper or lower boundary, given by the product of the the type (A) strands in either boundary (see Fig.~\ref{fig:bdy_op_update}c). This means that we can only form $n-2$ independent stabilizer generators in the canonical form with support on both upper and lower boundaries, leading to $\propto (n-2)$ entanglement between the two boundaries.

\section{Relation between Majorana partons and Jordan-Wigner fermions}\label{appsec:parton2jw}
The assignment of Majorana Fermions to spins can be achieved through two different methods, namely the Majorana Parton construction and the Jordan-Wigner transformation. In the former, four Majoranas are assigned to each spin, subject to an additional physical Hilbert space condition ($D_j = \gamma_{j,1} \gamma_{j,2} \gamma_{j,3} \gamma_{j,4}=1$), where operators that commute with $D_j$ are considered to be in the physical Hilbert space. In contrast, the Jordan-Wigner transformation assigns two Majorana Fermions to each spin. Our objective is to establish a correspondence between certain physical operators in the four Majorana representation and those in the two Majorana representation, when two of the four parton Majoranas are locked in a nearest-neighbor dimer state.

\begin{figure}
    \centering    \includegraphics[width= 0.8\columnwidth]{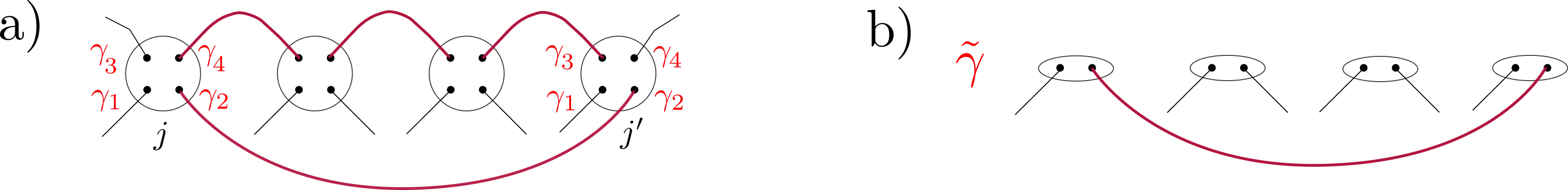}
    \caption{(a) A physical operator in terms of the parton Majoranas is represented (up to a sign) by the product of all the Majorana strands colored red in the figure. (b) The same operator can be represented as a bilinear operator of  Majoranas $\tilde{\gamma}$ obtained by Jordan-Wigner transformation of the qubits.  Thus, the Jordan-Wigner transformation of the qubit state is effectively the Majorana state on the $\gamma_{1,2}$ fermions.}
    \label{fig:phys_op}
\end{figure}

An example of a physical operator in the Majorana parton construction is given by Eq.~\ref{eq:phys_op} (see Fig.~\ref{fig:phys_op}a):

\begin{equation}
\begin{split}
O_{(j,s),(j',s')} = i\gamma_{j,s}\prod_{k=j'}^{k=j+1}(i\gamma_{k,3}\gamma_{k-1,4})\gamma_{j',s'}
\end{split}
\label{eq:phys_op}
\end{equation}
where $s$ and $s'$ are either 1 or 2. It can be shown that this operator commutes with all $D_j$'s because it has an even number of Majorana fermions labeled by any $k$. We can express these operators in terms of Pauli operators since they live in a physical Hilbert space. The relation between partons and Pauli operators is given by Eq.~\ref{eq:pauli_op}.

\begin{equation}
\begin{split}
    X_{j}= i\gamma_{j,1}\gamma_{j,2} = i\gamma_{j,4}\gamma_{j,3}  \\ Y_{j}=  i\gamma_{j,2}\gamma_{j,3} = i\gamma_{j,4}\gamma_{j,1} \\ Z_{j}= i\gamma_{j,1}\gamma_{j,3} = i\gamma_{j,2}\gamma_{j,4}, 
\end{split}
\label{eq:pauli_op}
\end{equation}

Using these relations, we can rewrite the physical operators in Eq.~\ref{eq:phys_op} in terms of Pauli operators, and an example is shown in Eq.~\ref{eq:jw_op}.

\begin{equation}
\begin{split}
O_{(j,1),(j',1)} = -Y_j\left(\prod_{k=j+1}^{k=j'-1}X_k\right)Z_{j'}
\end{split}
\label{eq:jw_op}
\end{equation}
Next, we use the Jordan-Wigner transformation to map these Pauli operators to Majorana operators by assigning two Majoranas to each spin. The mapping between Majorana operators and Pauli operators is given by Eq.~\ref{eq:jw}.

\begin{equation}
\begin{split}
\Tilde{\gamma}_{j,1} = \left(\prod_{k<j}X_k\right)Z_{j}, \ 
\Tilde{\gamma}_{j,2} = \left(\prod_{k<j}X_k\right)Y_{j}
\end{split}
\label{eq:jw}
\end{equation}

Using this transformation, we can rewrite the operators in Eq.~\ref{eq:jw_op} in terms of Majorana operators, as shown in Eq.~\ref{eq:jw2phys}.

\begin{equation}
\begin{split}
Y_j\left(\prod_{k=j+1}^{k=j'-1}X_k\right)Z_{j'} = -i\Tilde{\gamma}_{j,1}\Tilde{\gamma}_{j',1}
\end{split}
\label{eq:jw2phys}
\end{equation}

Using this mapping, any general physical operators in Eq.~\ref{eq:phys_op} can be represented as bilinears of $\tilde{\gamma}$ Majoranas,

\begin{equation}
\begin{split}
i\gamma_{j,s}\prod_{k=j'}^{k=j+1}(i\gamma_{k,3}\gamma_{k-1,4})\gamma_{j',s'} = i\Tilde{\gamma}_{j,s}\Tilde{\gamma}_{j',s'}
\end{split}
\label{eq:jw2parton}
\end{equation}

Here, $s$ and $s'$ are either 1 or 2. This shows that by removing $\gamma_3$ and $\gamma_4$ from any physical operator, we obtain the Jordan-Wigner transformed version of that operator, as shown in Fig.~\ref{fig:phys_op}b. Thus, the Jordan-Wigner transformation of the qubit state is effectively the Majorana state on the $\gamma_{1,2}$ fermions. Going the other direction, we can read out the qubit operator by performing a standard Jordan-Wigner transformation on the operator involving the $\gamma_{1,2}$ Majoranas only.  


\section{Jordan-Wigner transformation of the boundary after bulk measurements}\label{appsec:Gaussian}

In this section, we generalize the results from Appendix~\ref{appsec:parton2jw} from stabilizer states to general Gaussian states of Majorana operators.


\begin{figure}[ht]
    \centering    \includegraphics[width = 0.8\columnwidth]{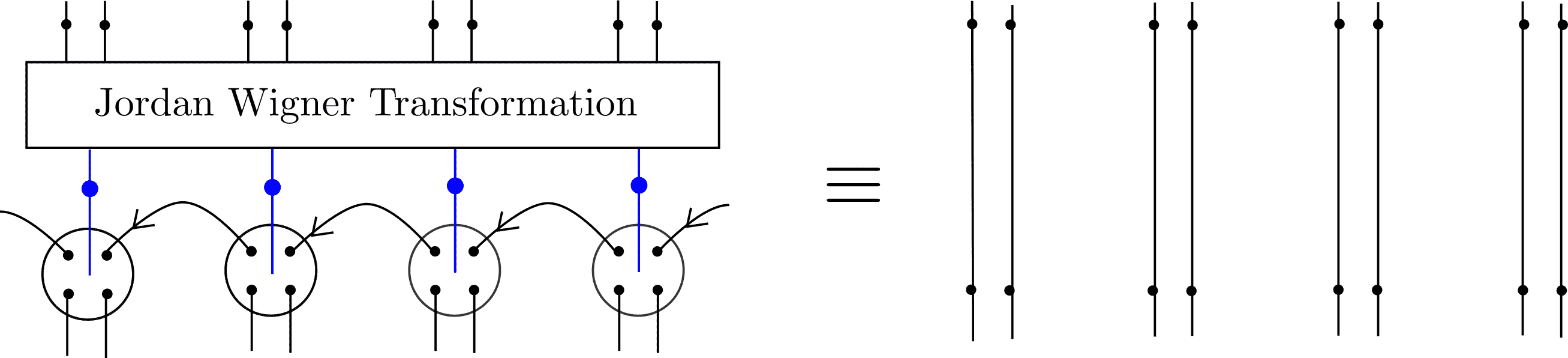} 
    \caption{The key result of this section is the cancellation between the projection tensor applied on the dimer state and the Jordan-Wigner transformation. Each large circle represents a projection tensor as defined in the paper. Two of the four Majoranas per circle are paired in the depicted nearest neighbor dimer state. By contracting these tensors in this manner, we obtain a tensor that maps $2N$ Majorana fermions (bottom) to $N$ spins (blue dots). Subsequently, the Jordan-Wigner tensor takes $N$ spins as input and outputs $2N$ Majorana fermions. These two tensors cancel each other out when combined.}
    \label{fig:jw_gaussian}
\end{figure}

We start with the density matrix of the final Gaussian state before projections and determine how the projections affect it. Any general Gaussian state can be represented by an orthogonal transformation of the original Majorana fermions $\gamma$, conveniently arranged as a vector $\Vec{\gamma} = (\gamma_{1,1},\gamma_{1,2},\gamma_{2,1}, ... , \gamma_{2,n})^{T}$. Let us define $\Vec{\gamma}'$ using a generic orthogonal matrix $O$ as $\Vec{\gamma}' = O \Vec{\gamma}$.

It is straightforward to show that $\tilde{\gamma}$ also satisfies the Majorana algebra. The most general boundary density matrix in our setup is composed of two parts: the first part comes from the initial short stabilizers on the boundary, and the second part comes from the Gaussian operations in the bulk. Such a general density matrix can be represented as~\cite{bravyi2004lagrangian}, 
\begin{eqnarray}
    \rho = \prod_{j=1}^{j=N-1} \frac{1+ i\gamma_{j+1,3}\gamma_{j,4}}{2} \prod_{k=1}^{k=N} \frac{1+ s_k (i\gamma'_{2k}\gamma'_{2k+1})}{2},
    \label{eq:gen_gaussianbdy}
\end{eqnarray}
where $s_{k} \in [-1,1]$. 

We need to apply the projections projections $P_{j} = \left(1+ \gamma_{j,1}\gamma_{j,2}\gamma_{j,3}\gamma_{j,4}\right)/2$ to the density matrix to get a physical density matrix,

\begin{eqnarray*}
    \rho_{f} = \left(\prod_{j=1}^{j=N} P_j\right)
    \;\rho
    \;\left(\prod_{j=1}^{j=N} P_j\right).
\end{eqnarray*}

We can re-express $i\gamma'_{2k}\gamma'_{2k+1}$ in terms of the original Majorana fermions as $\sum_{i,j} a^k_{i,j}i\gamma_{i}\gamma_{j}$, where $a^k_{i,j}=O_{2k,i}O_{2k+1,j}$. It is now straightforward to show that Eq.~\ref{eq:gen_gaussianbdy} can be rewritten as,

\begin{eqnarray}
    \rho = \prod_{j=1}^{j=N} \frac{1+ i\gamma_{j,3}\gamma_{j+1,4}}{2} \prod_{k=1}^{k=N} \frac{1+ s_k \sum_{i,j} a^k_{i,j}i\gamma_{i}(\prod_{k=j}^{k=i+1}(i\gamma_{k,3}\gamma_{k-1,4}))\gamma_{j}}{2},
\end{eqnarray}
where we have introduced a Majorana string in the second factor, which can always be absorbed in the first factor. 

We introduce a new notation to represent the density matrix as the product of two parts $\rho = \rho_{1}\rho_{2}$,
\begin{eqnarray*}
&\rho_1= \prod_{j=1}^{j=N} \frac{1+ i\gamma_{j,3}\gamma_{j+1,4}}{2}
\\
&\rho_2 =\prod_{k=1}^{k=N} \frac{1+ s_k \sum_{i,j} a^k_{i,j}i\gamma_{i}(\prod_{k=j}^{k=i+1}(i\gamma_{k,3}\gamma_{k-1,4}))\gamma_{j}}{2}.
\end{eqnarray*}

One can easily check that $\rho_2$ and $P_j$ commute, due to the fact that $\rho_2$ intersects even number of times in each factor of $P_j$. This allows us to represent the density matrix after projection as $\rho_{f} = \left(\prod P_{j}\right) \rho_{1} \left(\prod P_{j}\right) \rho_{2}$. The computation of $\left(\prod P_{j}\right) \rho_{1} \left(\prod P_{j}\right)$ is now amenable to stabilizer formalism. We can directly borrow the results of Appendix~\ref{appsec:parton2jw} to re-express the parton Majoranas in terms of the Jordan Wigner Majoranas, $\gamma \to \tilde{\gamma}$. As described in that Appendix, we can simply remove the $\gamma_{3}$ and $\gamma_{4}$ operators to represent a physical density matrix after projection in terms of the Jordan Wigner Majoranas $\tilde{\gamma}$,

\begin{eqnarray}
    \rho_{f} =  \prod_{k=1}^{k=N} \frac{1+ s_k \sum_{i,j} a^k_{i,j}i\tilde{\gamma}_{i}\tilde{\gamma}_{j}}{2}.
\end{eqnarray}

The $\tilde{\gamma}$ Majoranas can now be rotated using the orthogonal transformation $O$ to get back to the initial Gaussian state we began with in terms of $\tilde{\gamma}^{\prime}$ Majoranas,

\begin{eqnarray}
    \rho_{f} = \prod_{k=1}^{k=N} \frac{1+ s_k (i\tilde{\gamma}'_{2k}\tilde{\gamma}'_{2k+1})}{2}
\end{eqnarray}.

These series of manipulations can be depicted simply in Fig.~\ref{fig:jw_gaussian}. In summary, the boundary projections followed by a Jordan Wigner transformation lead us back to the original Gaussian state before enforcing the projections.

\end{document}